\documentclass[sigconf]{acmart}
\usepackage{amsmath,amsfonts,bm}

\def\eqref#1{equation~\ref{#1}}
\def\1{\bm{1}}

\DeclareMathAlphabet{\mathsfit}{\encodingdefault}{\sfdefault}{m}{sl}
\SetMathAlphabet{\mathsfit}{bold}{\encodingdefault}{\sfdefault}{bx}{n}

\newcommand{\sigmoid}{\sigma}

\usepackage[ruled,vlined]{algorithm2e}
\usepackage{import}
\usepackage{paralist}
\usepackage{subcaption}

\usepackage{caption}
\usepackage{multirow}
\usepackage{makecell}
\usepackage{threeparttable}
\usepackage{graphicx} 
\AtBeginDocument{%
  \providecommand\BibTeX{{%
    \normalfont B\kern-0.5em{\scshape i\kern-0.25em b}\kern-0.8em\TeX}}}

\copyrightyear{2022}
\acmYear{2022}
\setcopyright{acmcopyright}\acmConference[WSDM '22]{Proceedings of the Fifteenth
ACM International Conference on Web Search and Data Mining}{February 21--25,
2022}{Tempe, AZ, USA}
\acmBooktitle{Proceedings of the Fifteenth ACM International Conference on Web
Search and Data Mining (WSDM '22), February 21--25, 2022, Tempe, AZ, USA}
\acmPrice{15.00}
\acmDOI{10.1145/3488560.3498505}
\acmISBN{978-1-4503-9132-0/22/02}
\graphicspath{./img}

\begin{document}

\fancyhead{}
\title{Heterogeneous Global Graph Neural Networks for Personalized Session-based Recommendation }


% \author[1]{Yitong Pang}

% \author[2]{Lingfei Wu}

% \author[1]{Qi Shen}

% \author[1]{Yiming Zhang}

% \author[1]{Zhihua Wei}
% \authornote{Corresponding author.}

% \affil[1]{Department of Computer Science and Technology, Tongji University}
% \affil[2]{JD Silicon Valley Research Center}

% old version
\author{Yitong Pang}
\authornote{Both authors contributed equally to this research.}
\affiliation{%
  \institution{Tongji University}
    \country{China}
  }
\email{1930796@tongji.edu.cn}

\author{Lingfei Wu}
\authornotemark[1]
\affiliation{%
  \institution{JD.COM Silicon Valley Research Center}
   \country{USA}
  }
\email{lwu@email.wm.edu}

\author{Qi Shen}
\affiliation{%
  \institution{Tongji University}
      \country{China}
  }
\email{1653282@tongji.edu.cn}

\author{Yiming Zhang}
\affiliation{%
  \institution{Tongji University}
      \country{China}
      }
\email{2030796@tongji.edu.cn}

\author{Zhihua Wei}
\authornote{Corresponding author.}
\affiliation{%
  \institution{Tongji University}
      \country{China}
      }
\email{zhihua_wei@tongji.edu.cn}

\author{Fangli Xu}
\affiliation{%
  \institution{Squirrel AI Learning}
      \country{USA}}
\email{fxu02@email.wm.edu}

\author{Ethan Chang}
\affiliation{%
  \institution{Middlesex School}
        \country{USA}}
\email{echang@mxschool.edu}

\author{Bo Long}
\affiliation{
  \institution{JD.com}
      \country{China}}
\email{bo.long@jd.com}

\author{Jian Pei}
\affiliation{
  \institution{Simon Fraser University}
      \country{Canada}}
\email{jpei@cs.sfu.ca}

\renewcommand{\shortauthors}{XXX, et al.}

\begin{abstract}
Predicting the next interaction of a short-term interaction session is a challenging task in session-based recommendation. Almost all existing works rely on item transition patterns, and neglect user historical sessions while modeling user preference, which often leads to non-personalized recommendation. And existing personalized session-based recommenders are limited to sessions of the current user, and ignore the useful item-transition patterns from other user’s historical sessions. To address these issues, we propose a novel Heterogeneous Global Graph Neural Networks (HG-GNN) to exploit the item transitions over all sessions in a subtle manner for better inferring user preference from the current and historical sessions. To effectively exploit the item transitions over all sessions from users, our global graph contains item transitions of sessions, user-item interactions and global co-occurrence items. Moreover, to capture user preference from sessions comprehensively, we propose a graph augmented preference encoder to learn the session representation. Specifically, we design a novel heterogeneous graph neural network (HGNN) on heterogeneous global graph to learn long-term user preference and item representations with rich semantics. Based on the HGNN, we propose the Personalized Session Encoder to combine the general user preference and temporal interest of the current session to generate the personalized session representation for recommendation.  Extensive experimental results on three real-world datasets show that our model outperforms other state-of-the-art methods. The implementation of our proposed model is publicly
available at \textit{https://github.com/0215Arthur/HG-GNN}.

\end{abstract}

\begin{CCSXML}
<ccs2012>
<concept>
<concept_id>10002951.10003317.10003347.10003350</concept_id>
<concept_desc>Information systems~Recommender systems</concept_desc>
<concept_significance>500</concept_significance>
</concept>
</ccs2012>
\end{CCSXML}

\ccsdesc[500]{Information systems~Recommender systems}

\keywords{Recommendation system; Session-based recommendation; Graph neural network}


\maketitle

\section{Introduction}

Recommendation systems are widely used in online platforms, as an effective tool for addressing information overload. Recently, in some real-world applications (e.g. stream media), recommendation systems need to focus on the interactions within the active session. However, traditional recommendation methods (e.g. collaborative filtering \citep{sarwar2001item}) that usually learn user preferences from the long-term  historical interactions, are typically not suitable for these scenarios.
Therefore, session-based recommendation has attracted great attention in the past few years, which generates recommendations mainly using interactions in the active session.

The current mainstream methods for session-based recommendation are modeling the sequential patterns of item transitions within the current session. Many deep learning-based models have been proposed for session-based recommendation, which utilize the item transition sequences to learn the representation of the target session \citep{gru4rec,narm,srgnn,fgnn,lessr,gce-gnn}. For instance, several methods rely on the sequence modeling capability of recurrent neural networks (RNNs) to extract the sequential feature of the current session \citep{gru4rec,narm}. %More recently, a few works attempt to convert the session sequences to graphs and infer the preferences for a given session through the graph neural networks \citep{srgnn,fgnn,lessr,xu2019graph,gce-gnn}.

\begin{figure}[t]
    \centering
    \includegraphics[width=\linewidth]{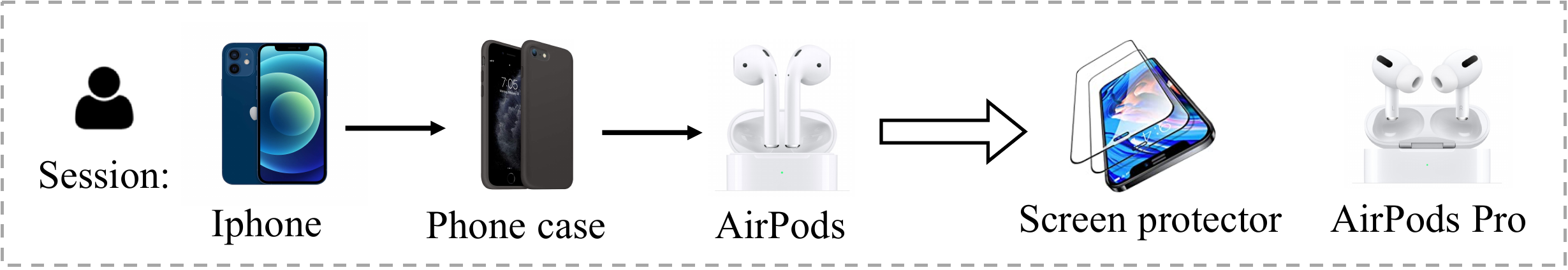}
    \caption{A toy example of session-based recommendation.}
    \label{fig:exmaple}\vspace{-0.6cm}
\end{figure}

Although these approaches have achieved encouraging results, they still suffer several significant limitations. First, most methods fail to consider user historical sessions, as users are assumed to be anonymous, leading to non-personalized recommendations. These methods mainly rely on the item information of the current session to form the session representation, and then utilize the item correlation to make recommendations, while the influence of user features and historical behaviors are overlooked. For instance, given user A and B, they have the similar session sequences, i.e. \emph{Iphone} $\rightarrow$ \emph{Phone case} $\rightarrow$ \emph{AirPods}. As shown in \autoref{fig:exmaple}, these previous methods usually  generate the same candidate items for different users: \emph{Screen protector}, \emph{AirPods Pro}, etc. It is worth to note that in many cases, the user may be logged in or some form of user identifier (cookie or other identifier) may exist. In these cases it is reasonable to assume that user behaviors in historical sessions are useful when providing personalized recommendations. 

Second, the existing studies on personalized session-based recommendation model user preference only based on the session,  while the useful item-transition patterns from other user’s historical sessions are ignored. These approaches capture the impact of historical sessions on the current session of the user to obtain the personalized session representation via the RNN-based or GNN-based models \citep{H-RNN, DANN, A-PGNN, chen2021efficient}. Conceptually, utilizing the item-transition of the other sessions can capture more complex item correlations and might help model the user preference.

To address these aforementioned limitations and achieve personalized session-based recommendation, we propose a novel approach to exploit the item transitions over all sessions in a subtle manner for better inferring the user preference from the current and historical sessions, named Heterogeneous Global Graph Neural Networks (HG-GNN). 
To effectively exploit the item transitions over all sessions from users, we first propose a novel heterogeneous global graph consisting of user nodes and item nodes. In particular, we utilize user-item historical interactions to construct user-item edges in the graph in order to capture long-term user preference. Then we adopt pairwise item-transitions in the session sequence to construct connections between items. To capture the potential correlations, we calculate similar item pairs based on the global co-occurrence information to construct item edges.
Moreover, to capture user preference from sessions comprehensively,  we propose to learn preference representations from the heterogeneous global graph via a graph augmented preference encoders. 
We propose a novel heterogeneous graph neural network (HGNN) on the heterogeneous global graph to learn the long-term user preference and item representations with rich semantics.
Furthermore, Personalized Session Encoder combines the item information of the current session and the general user preference to generate the personalized session representation, which is used to generate the more accurate and personalized recommendation list.

\begin{figure*}
    \centering
    % \includegraphics[width=\linewidth]{img/overview.pdf}
    % \scriptsize
    % \def\svgwidth{0.7\linewidth}
    % \import{img/}{framework.png}
    % \includegraphics[width=0.6\textwidth]{img/hgnn_v4.pdf}
    \includegraphics[width=0.6\textwidth]{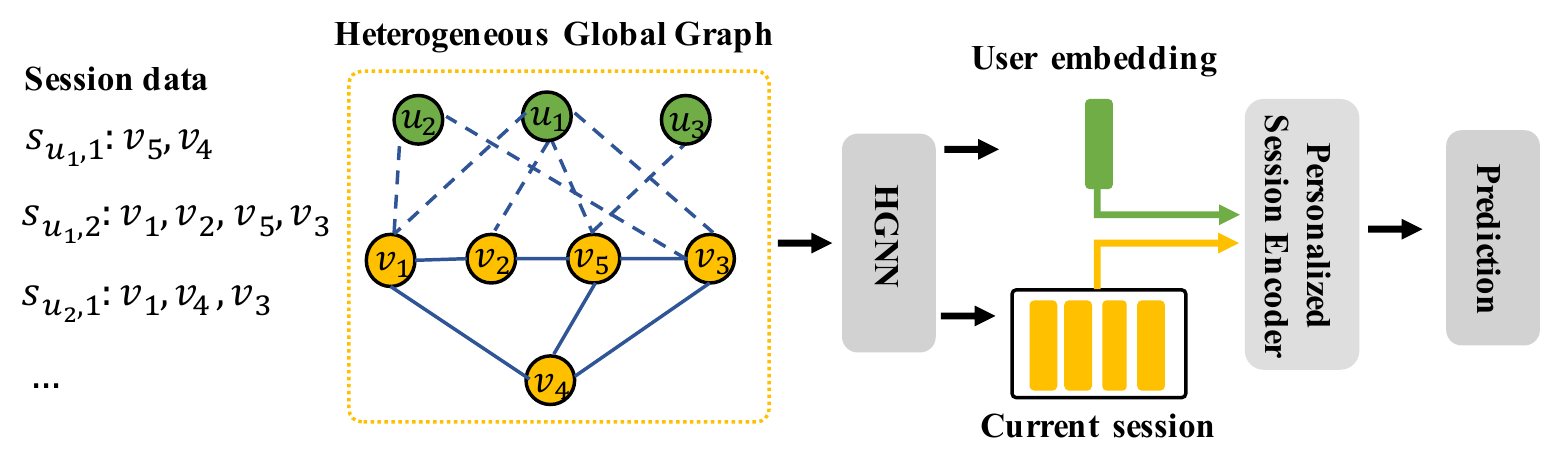}
    
    \caption{The overview of HG-GNN.}
    \label{fig:framework}\vspace{-0.3cm}
\end{figure*}
% The heterogeneous graph  contains two types of meta-paths: \emph{user to item} and \emph{item to item}.
% The overview of HG-GNN. The heterogeneous global graph  contains two types of meta-paths: \emph{user to item} and  \emph{item to item}. The global graph is passed as input to the HGNN layer to learn the user and  item embeddings with rich semantics. We adopt the User Preference Attention to learn the impact of long-term user preference on current session. Meanwhile, we employ the Dynamic Interest Learning and General Interest Learning modules to comprehensively capture the current session preference. Finally, we aggregate the historical and current preference representations to generate the final user embedding.

Our main contributions of this work are summarized below:
\begin{itemize}
    \item We propose a novel heterogeneous global graph to effectively exploit item transitions over sessions, which consists of item transitions, user-item interactions, and similar pairs constructed based on global co-occurrence information. 
    \item We propose a graph augmented hybrid encoder which consists of a heterogeneous graph neural network and a personalized session encoder to generate the session preference embedding for personalized session-based recommendation.
    \item Extensive experiments on three datasets demonstrate that our model is superior compared with state-of-the-art models.
\end{itemize}

\vspace{-0.3cm}
\section{Related Works}

Next we review related works of session-based recommendations.

\textbf{Traditional Methods.} In early research, the session-based recommender systems were mainly based on co-occurrence information following collaborative filtering \citep{sarwar2001item,dias2013improving,wang2016spore}.  To capture the sequential patterns in the session, Markov chain-based methods predicted the next action of users given the last action \citep{rendle2010factorizing,shani2005mdp}. For instance, FPMC \citep{rendle2010factorizing} factorized personalized transition matrices by a matrix factorization based first-order Markov chain. However, Markov chain-based methods usually only model first-order transitions and cannot capture more complex sequential patterns.

\textbf{Deep Learning-based Methods.} Recently, deep learning-based methods have been widely used for session-based recommendations, including RNN-based and CNN-based methods \citep{gru4rec,narm,liu2018stamp,ren2019repeatnet,song2019islf}.  GRU4Rec \citep{gru4rec} was the first RNN-based method for session-based recommendation, and it captured item interaction sequences by GRU layers. NARM \citep{narm} employed the attention mechanism with RNN to learn the more representative item-transition information for session-based recommendation, an approach which has been proved to be effective  for learning session representations. Caser \citep{2018Personalized} represented the session items  with latent matrix and adopted CNN to learn  general preferences and sequential patterns.

% Many subsequent RNN-based methods  liu2018stamp,  utilized the last click information to generate session embedding by incorporating attention mechanism.

\textbf{GNN-based Methods.} 
Recent years have seen a surge of interests in Graph Neural Networks \citep{hamilton2017sage,kipf2016semi,xu2018graph2seq,chen2019reinforcement,chen2020iterative} and as a result various GNN methods have been utilized for improving recommendation system. 
Several GNN-based models have been proposed to learn item representations for the session-based recommendation \citep{srgnn,xu2019graph,fgnn,wang2020beyond,lessr,gce-gnn, 2020DHCN}. For example, SR-GNN \citep{srgnn}  applied the gated GNN (GGNN) \citep{li2015gated} to a directed session graph to learn item embeddings. Based on SR-GNN,  GC-SAN \citep{xu2019graph}  was proposed, a model that adopted self-attention mechanism to capture global dependencies between different positions and integrated GGNN to generate the session embeddings. \citep{wang2020beyond} proposed to construct a multi-relational item graph for session-based behavior prediction. Chen et al. \citep{lessr} proposed a lossless encoding scheme that preserved the edge-order for better modeling of the conversion of sessions to graphs. Additionally, it proposed a shortcut graph attention layer to capture long-range dependencies between items in the session.  GCE-GNN \citep{gce-gnn} proposed a framework that adopted GNNs to learn session-level and global-level item embeddings from the directed session graph and the undirected global item graph, and integrated the different level  item embeddings to generate the final session embedding. However, these methods fail to consider the historical user sessions, and mainly focus on the item sequence in the session, leading to non-personalized recommendations.

% DHCN \citep{2020DHCN}  transformed the session data to a hyper-graph to model the high-order correlations among items, and employed graph convolution in both the global hyper-graph and the line graph between different sessions to improve recommendation performance. 

\textbf{Personalized Session-based Recommendation.} The existing research on the personalized session-based recommendation is still in the early stage. Existing methods mainly focus on the use of the user’s historical session, which also called session-aware recommendations \citep{HTCN,H-RNN,DANN,A-PGNN, latifi2021session}. H-RNN \citep{H-RNN} is the most representative RNN-based approaches for personalized session-based recommendation, which utilized a hierarchical RNN to capture users’ short- and long-term preferences from the historical sessions of the current user. Recently, A-PGNN \citep{A-PGNN} converted the each user behaviors into graphs, and extracted the personalized structural information via the GNN. For the social recommendation task, \citep{chen2021efficient} proposed a framework that combined the social network and user-item interactions to form a global knowledge graph and exploited to learn the user and item embeddings via the GNNs from the graph. This framework is not effectively designed for session data and is not suitable for more general session-based  recommendations.

These methods only use the sessions of current user and ignore useful item-transition patterns from other user’s historical sessions. Historical sessions of other users can provide useful item transition patterns and might help model the current user preference.

\section{Preliminary}\label{sec:space}
% \subsection{Neural Architecture Search Spaces}\label{sec:space}

In existing literature, next-item recommendation is a sub-area of session-based recommendation \citep{wang2021survey}. It aims to predict items that the user will click next, based on historically interacted items of the active session. The current mainstream research mainly focuses on learning the patterns of item transitions in the current session, which usually ignores the past sessions of the user. To achieve accurate and personalized recommendation, we consider the user information of the session in this paper. We present the formulation of the problem researched in this paper as below.

Let $V=\{v_1,v_2,...,v_{|V|}\}$ and $U=\{u_1,u_2,...,u_{|U|}\}$ denote the universal set of items and users, respectively. We denote  the interaction records of each user as $S_{u_i}=\{S_{u_{i,1}},S_{u_{i,2}},...,S_{u_{i,n}}\}$, which contain the historical sessions of user $u_i$ in chronological order, where $S_{u_{i,j}}$ is the $j$-th session sequence of $u_i$.  The session $S_{u_{i,j}}=\{v_1^{u_{i,j}},v_2^{u_{i,j}},...,v_l^{u_{i,j}}\}$ is a sequence of items in chronological order, where $v_t^{u_{i,j}} \in V$ represents the item which interacted with user $u_i$ at time step $t$, and $l$ denotes the session length.

Given the historical sessions $\{S_{u_{i,1}},S_{u_{i,2}},...,S_{u_{i,t-1}}\}$ and current session $S_{u_{i,t}}$ of $u_i$, the goal of  personalized session-based recommendation is to predict the next item $v_{l+1}^{u_{i,t}}$ that user $u_i$ is most likely to click. Specifically, the recommendation model is trained to generate the probability score for each candidate, i.e., $\hat{\mathbf{y}}=\{\hat{y}_{1},\hat{y}_{2},...,\hat{y}_{{|V|}}\}$ where $\hat{y}_i$  denotes the prediction score of item $v_i$.

\section{Methodology}\label{sec:model}
\subsection{Overview}

In this section, we detail the design of our model, Heterogeneous Global Graph Neural Network for personalized session-based recommendation (HG-GNN). As shown in \autoref{fig:framework}, to effectively exploit the item transitions over all sessions from users, we propose a novel heterogeneous global graph to organize historical sessions and exact the global information. We utilize user-item historical interactions edges to construct user-item edges in the graph in order to capture user preference. To utilize the potential correlations, we calculate similar item pairs based on the global co-occurrence information to construct item edges. The pairwise item-transitions in the session sequence are also used to construct connections.

Furthermore, we propose to learn session representation from the heterogeneous global graph via a graph augmented session preference encoders. Specifically, we propose a novel heterogeneous graph neural network on the heterogeneous global graph to learn the long-term user preference and item representations with rich semantics.
Moreover, we employ the Personalized Session Encoder which combines the user general preference and items of the current session to generate personalized session representation comprehensively. Next, we will introduce these modules in detail.

% Next, we employ the Current Preference  Encoder which consists of the dynamic and general interest learning modules to comprehensively capture the current preference from the user current session. Meanwhile, in the Historical Preference Encoder, we utilize the user preference attention to learn the impact of long-term user preference on the current session. To make the personalized session-based recommendation, we combine the representations of the user current preference and the historical interests to generate the final user embedding. We will introduce these modules in detail in next subsections.

\subsection{Heterogeneous Global Graph Construction}
In this subsection, we mainly describe how to construct a  heterogeneous global graph with two meta-paths:  \emph{item-to-user} and \emph{item-to-item}. In this paper, we transform all training session sequences into a directed heterogeneous global graph $\mathcal{G}=(\mathcal{V},\mathcal{E})$, where nodes are $\mathcal{V}$ which consists of user nodes $u_i \in U$ and item nodes $v_i \in V$. The edges in $\mathcal{E}$ contain the different relations of meta-paths, and each edge is denoted as $(v_{i},v_{j},r)$.

\subsubsection{Item-to-Item.}

The item transition information is the basis for session-based recommendations. The transition relationship between items can include the adjacent interaction behaviors in the session and frequent co-occurrence behaviors between the same session. The two behaviors are complementary to each other. Different from the existing methods \citep{lessr,srgnn,gce-gnn} that only use the adjacent interaction relationship in the session to build the session graph or the sequence model, we adopt both types of information to construct the global graph.

Similar to \citep{srgnn,lessr,gce-gnn}, we define two edges $(v_i,v_j,r_{in})$,$(v_j,v_i,r_{out})$ for the transition from $v_i$ to $v_j$ in the session. Meanwhile, for each item node $v_i$, we generate weight for
its adjacent edges to distinguish the importance of $v_i$'s neighbors
as follows: For each edge ($v_i$, $v_j$, $r$) where ($r \in \{r_{in}, r_{out}\}$), we utilize its frequency over all the historical sessions as the edge weight. To ensure the relevance of items, we only sample the top-$S$ edges with the highest weights for each item $v_i$ on $\mathcal{G}$.

Additionally, we adopt the co-occurrence information to construct the edges between items. The frequent co-occurrence behaviors of two items in different sessions can show the strong item correlations. Specifically, for item $v_i$, we calculate its co-occurring items based on all historical sessions, and we select top-$k$ items that co-occur with $v_i$ frequently. Thus we define the edge $(v_j,v_i,r_{similar})$ for the two co-occurrence items $v_i$ and $v_j$.  The co-occurrence frequency between item $v_i$ and $v_j$ is calculated as followed:
% This relation is always more reliable than the adjacent interaction in the same session.

\begin{equation}
    f_{v}(v_i,v_j)=\frac{\sum_{s \in N(v_i) \cap N(v_j)} \frac{1}{|N(s)|}}{\sqrt{|N(v_i)||N(v_j)|}}\,,
\end{equation}
where $N(v_i)$ represents the set of sessions in which $v_i$ occurs. There may be two edges  $r_{in}$ and $r_{similar}$ between two items at the same time. In this case, we only keep the former edge. To avoid introducing too much noise, for each item node $v_i$, we use the number of adjacent interaction items of $v_i$ to cut off the top-$K$. Hence, only $K'_{v_i}$ edges are built as $r_{similar}$ for item node $v_i$:
\begin{equation}
    K_{v_i}' = min\{K, |N_{v_i}|\}\,, 
\end{equation}
\begin{equation}
    N_{v_i} = \{v_j | (v_i, v_j, r)\ where \  r \in \{r_{in}, r_{out}\}\}\,.
\end{equation}

\subsubsection{Item-to-User.}
The item-to-user meta-path directly represents the interaction behavior between the user and the item, revealing the implicit long-term preferences of the user. We convert the user-item interaction  into two types of directed edges in the graph $\mathcal{G}$:  $(v_j,u_i,r_{interact})$ and  $(u_i,v_j,r_{interacted\_by})$, which denote that user $u_i$ has interacted with item $v_j$. %According to the definition above, user-item interactions can be directly added to the graph. %However, it is not profitable to use all interactions without distinction for modeling user preferences. Therefore, we follow the sampling strategy to select representative interaction targets for users and items respectively. Specifically, for user $u_i$, we select top-$k$ popular items and top-$k$ cold items that they have interacted with to form edges. While for item $v_i$, the top-$k$ active users and top-$k$ cold users who have interacted with them are selected. User active score and item popularity are calculated by the number of user interactions and item appearance frequency, respectively.

In summary, we construct a novel heterogeneous global graph $\mathcal{G}$ with two types of nodes to effectively organize the session data. This global graph contains the basic pairwise item-transitions in sessions, the user-item historical interactions and the global co-occurrence information. We can unify the learning of user and item representations via the global graph, and capture item correlation and the long-term user preference. It is worth noting that other extra attributes of users or items can be easily integrated to construct extra edges or enhance node representation.

\subsection{Heterogeneous Global Graph Neural Network}
% In this subsection, we present the details of the HGNN layer on the directed  global heterogeneous graph . $\mathcal{G}$ contains item-transition information from all training sessions, and also encodes the user-item interaction records. 
We propose a heterogeneous graph neural network (HGNN) on the directed heterogeneous global graph $\mathcal{G}=(\mathcal{V},\mathcal{E})$ to encode the user and item representations, inspired by \citep{hamilton2017sage,schlichtkrull2018modeling}. The item IDs and user IDs are embedded in $d$-dimensional space and are used as initial node features in our model, $\mathbf{q}_{u_i}^{(0)}\in \mathbb{R}^d$ and $\mathbf{p}_{v_j}^{(0)}\in \mathbb{R}^d$ . Let $\mathbf{q}_{u_i}^{(k)}$ and $\mathbf{p}_{v_j}^{(k)}$ denote the refined embedding of user $u_i$ and item $v_j$ after $k$ layers propagation.

% Such a process can be described as:  
% \begin{equation}
%     \mathbf{x}_i^{k+1}=\text{AGG}(\mathbf{x}_i^{(k)},\{x_j^{(k)} \in \mathcal{N}_i\})\,,
% \end{equation}
% where  $\text{AGG}(\cdot)$ is the aggregation function and $\mathcal{N}_i$ denotes the neighbors of node $i$. 
 
In the GNN layer, the node representations are updated by aggregating features of neighbors and passing messages along edges.  According to the definition of the directed heterogeneous graph in the previous subsection, for the item node, there exist three types of edges connecting the item neighbors i.e., $r_{in}$, $r_{out}$ and $r_{similar}$. There is one type of edge connecting the user neighbors: $r_{interacted\_by}$.  For each specific edge-type $r_x$, we accumulate messages over all neighbors $\mathcal{N}_{r_x}(v_j)$ . The aggregation process can be denoted as follows:

\begin{equation}
\begin{aligned}
    \mathbf{p}^{(k+1)}_{\mathcal{N}_{r_x}(v_i)}=\frac{1}{|\mathcal{N}_{r_x}(v_i)|}\sum_{n\in \mathcal{N}_{r_x}(v_i)}\mathbf{e}^{(k)}_{n},
\end{aligned}\label{eqn:grad}
\end{equation}
\begin{equation}
\begin{aligned}
    \mathbf{p}^{(k+1)}_{v_i,r_x}=f(\mathbf{W}^{(k+1)}_{r_x}[\mathbf{p}^{(k+1)}_{\mathcal{N}_{r_x}(v_i)}||\mathbf{p}^{(k)}_{v_i}]+\mathbf{b}_{r_x}^{(k+1)})\,,
\end{aligned}\label{eqn:grad}
\end{equation}
where $\mathcal{N}_{r_x}(v_j)$ denotes the neighbors of $v_j$ with the edge type $r_x$, $\mathbf{W}_{r_x}^{(k+1)} \in \mathbb{R}^{d \times d}$, $||$ indicates concatenation operation and $\mathbf{b}_{r_x}^{(k+1)} \in \mathbb{R}^{d}$ are edge-type specific parameters. $f(\cdot)$ denotes the activation function, and we choose the relu function for activation. $\mathbf{e}_n^{(k)}$ is the representation of neighbor node in the $k$-th layer, i.e.,  $\mathbf{e}_n^{(k)}=\mathbf{p}_n^{(k)}$ or $\mathbf{e}_n^{(k)}=\mathbf{q}_n^{(k)}$.

For item $v_i$, we accumulate different messages propagated by different types of edges and update the item node representation, as shown in the following formula:
\begin{equation}
\begin{aligned}
    \mathbf{p}^{(k+1)}_{v_i}=\text{accum}(\mathbf{p}^{(k+1)}_{v_i,r_u},{\mathbf{p}^{(k+1)}_{v_i,r_{in}}},{\mathbf{p}^{(k+1)}_{v_i,r_{out}}},{\mathbf{p}^{(k+1)}_{v_i,r_{similar}}}),
\end{aligned}\label{eqn:grad}
\end{equation}
where $\text{accum}(\cdot)$ denotes an accumulation operation, such as $\text{sum}(\cdot)$ or $\text{stack}(\cdot)$. $r_u$ is the short for $r_{interacted\_by}$. In practice, we adopt the $\text{mean}(\cdot)$ operation to take the average of all messages.

The aggregation operation of the user node is similar to the item operation as above. The updated user node representation is as follows:
\begin{equation}
\begin{aligned}
    \mathbf{q}^{(k+1)}_{u_i}=\text{accum}(\mathbf{q}^{(k+1)}_{u_i,r_v})\,,
\end{aligned}\label{eqn:grad}
\end{equation}
where $r_v$ is the short for $r_{interact}$.

After $K$ layers of HGNN, we combine the embeddings obtained at each layer to form the final global-level representation of a user and item:
\begin{equation}
\begin{aligned} 
    \mathbf{q}_{u_i}=\sum_{k=0}^K{\alpha_{k}}\mathbf{q}_{u_i}^{(k)};\quad \mathbf{p}_{v_i}=\sum_{k=0}^K{\alpha_{k}}\mathbf{p}_{v_i}^{(k)}, 
\end{aligned}\label{eqn:grad}
\end{equation}
where $\alpha_k \geq 0$ indicates the importance of the $k$-th layer output. In the experiment, we empirically set $\alpha_k$ uniformly as $1/(K+1)$ following \citep{he2020lightgcn}. Through the HGNN layer, we can learn the long-term user preference and  global-level refined item embeddings which contain rich semantics.

% Current Preference Encoder
\subsection{Personalized Session Encoder}\label{sec:search}
Modeling user preferences based on the current session requires consideration of the drifted preference and general preference of the user. Conceptually, user preferences are always dynamic and can be learned from the current session. Thus \emph{Current Preference Learning} module is proposed to obtain the item transitions pattern and model the sequential feature of the ongoing session.

Meanwhile, the user's next interaction behavior is also flexible and can be influenced by the long-term stable user preference. Therefore, we propose a \emph{General Preference Learning} module to capture the general preference based on the attention mechanism.

\subsubsection{Current Preference Learning}
% Previous works \citep{srgnn,narm,lessr,gce-gnn} show that the item sequence of the session is essential to the session-based recommendation. In this module, we mainly adopt the RNN-based methods to capture the dynamic interest from the item sequence. Given the current session $s$, we select the last $M$ items $\{v_1^s,v_2^s,...,v_M^s\}$ from $s$ as the input of DIL module. Firstly, we employ the GRU to capture the sequential feature of the current session as follows:

% In this part, we employ a position-aware attention mechanism to capture the general user interest of the current session.  
% We employ the attention mechanism to learn the importance of items in current session, and perform feature aggregation to obtain the general representation of the current session. In essence, GIL can capture the  basic item correlations and general feature of the user interactions in the current session. We also consider the influence of the item position in the modeling.

% \begin{equation}
% \begin{aligned}
%     \mathbf{h}_{t}^s=\text{GRU}(\mathbf{p}_{v_1^s}^{(0)},\mathbf{p}_{v_2^s}^{(0)},...,\mathbf{p}_{v_t^s}^{(0)})
% \end{aligned}\label{eqn:gru}
% \end{equation}
% where $\mathbf{h}_{t}^s$ denotes the $t$-th item sequential vector of the session $s$. 

To capture the user’s main purpose in the current session and represent the current session as an embedding vector, we apply an item-level attention mechanism which dynamically selects and linearly combines different item information of the current session following \citep{narm, srgnn, lessr}. The contribution of each item to the session representation is often influenced by the item position information (i.e., the chronological order in the session sequence) \citep{gce-gnn}. Therefore, given the last $l$ items of the current session $\{v_1^s, v_2^s, ..., v_l^s\}$, we concatenate the item representation $\mathbf{p}_{v_i}$ learnt from the HGNN layer with the reversed position embedding as follows :
\begin{equation}
\begin{aligned}
    \mathbf{p}_{v_i^s}'=\mathbf{W}_c[\mathbf{p}_{v_i^s}|| \mathbf{l}_{i}]\,,
\end{aligned}\label{eqn:graph}
\end{equation}
where position embedding $\mathbf{l}_{i} \in \mathbb{R}^d$ and $\mathbf{W}_{c} \in \mathbb{R}^{d \times 2d}$ are trainable. 

The basic session preference can be denoted as the average of item representations of the session: 
\begin{equation}
\begin{aligned}
    \mathbf{p}_s'=\frac{1}{l}\sum_{i=1}^l\mathbf{p}_{v_i^s}'\,,
\end{aligned}\label{eqn:graph}
\end{equation}
Moreover, we adopt the soft-attention mechanism to learn the corresponding weight of each item in the session:
\begin{equation}
\begin{aligned}
    \alpha_i =\text{softmax}_i(\epsilon_i),
\end{aligned}\label{eqn:graph}
\end{equation}
\begin{equation}
\begin{aligned}
    \epsilon_{i}=\mathbf{v}_0^T\sigmoid(\mathbf{W}_0\mathbf{p}_{v_i^s}'+\mathbf{W}_1\mathbf{p}_s'+\mathbf{b}_0)\,,
\end{aligned}\label{eqn:graph}
\end{equation}
where $\sigmoid(\cdot)$ is the sigmoid function, $\mathbf{W}_0\,, \mathbf{W}_1 \in \mathbb{R}^{d \times d}$ and $\mathbf{v}_0\,,\mathbf{b}_0 \in \mathbb{R}^d$ are trainable parameters. Finally, the current preference   $\mathbf{C}_u$  of the given session can be denoted as follows:
\begin{equation}
\begin{aligned}
    \mathbf{C}_u =\sum_{i=1}^l\alpha_i\mathbf{p}_{v_i^s}'\,.
\end{aligned}\label{eqn:graph}
\end{equation}
% The current preference of the current session $\mathbf{C} \in \mathbb{R}^{d}$ is formed by the last $l$ items, where 

\subsubsection{General Preference Learning}
The user embedding $\mathbf{q}_u$ learnt from the HGNN layer contains the long-term stable preference of user $u$. We consider the correlation between items in the current session  and the user general preference, and learn the corresponding weights through the attention mechanism: 
\begin{equation}
\begin{aligned}
    \gamma_i =\text{softmax}_i(e_i),
\end{aligned}\label{eqn:graph}
\end{equation}
\begin{equation}
\begin{aligned}
    e_i=\mathbf{v}_1^T\sigmoid(\mathbf{W}_2\mathbf{p}_{v_i^s}+\mathbf{W}_3\mathbf{q}_{u}+\mathbf{b}_1)\,,
\end{aligned}\label{eqn:graph}
\end{equation}
where $\mathbf{W}_2\,, \mathbf{W}_3 \in \mathbb{R}^{d \times d}$ and $\mathbf{v}_1\,,\mathbf{b}_1\in \mathbb{R}^d$ are trainable parameters.
Finally, we linearly combining the item representations to obtain the general preference of the current session:
\begin{equation}
\begin{aligned}
    \mathbf{O}_u =\sum_{i=1}^l\gamma_i\mathbf{p}_{v_i^s}\,.
\end{aligned}\label{eqn:graph}
\end{equation}
% \begin{equation}
% \begin{aligned}
%     \mathbf{g}_{u}^s=\sum_{i=1}^M\beta_i\mathbf{m}_{v_i}\,,
% \end{aligned}\label{eqn:graph}
% \end{equation}
% \begin{equation}
% \begin{aligned}
%     \beta_i=\mathbf{r}^T\sigmoid(\mathbf{W}_2\mathbf{m}_{v_i^s}+\mathbf{W}_3\mathbf{s}'+\mathbf{b_{s}})\,,
% \end{aligned}\label{eqn:graph}
% \end{equation}
% \begin{equation}
% \begin{aligned}
%     \mathbf{s}'=\frac{1}{M}\sum_{i=1}^M\mathbf{m}_{v_i^s}\,.
% \end{aligned}\label{eqn:graph}
% \end{equation}
% where $\mathbf{W}_2\,, \mathbf{W}_3 \in \mathbb{R}^{d \times d}$ and $\mathbf{r}\,,\mathbf{b}_{s}\in \mathbf{R}^d$ are trainable parameters.

% Finally, we use the gate mechanism to combine user long-term preferences and historical fusion representation $\mathbf{s}_u^s$, to get the historical preference representation $\mathbf{o}_{u}^h$: 
% \begin{equation}
% \begin{aligned}
%     \alpha_{u}=\sigmoid(\mathbf{W}_c(\mathbf{q}_{u} || \mathbf{s}_{u}^s))\,,
% \end{aligned}\label{eqn:graph}
% \end{equation}
% \begin{equation}
% \begin{aligned}
%     \mathbf{o}_{u}^h=\alpha_{u}\cdot \mathbf{q}_{u} + (1 - \alpha_{u})\cdot \mathbf{s}_{u}^s)\,.
% \end{aligned}\label{eqn:graph}
% \end{equation}

We argue that the two different types of preference representations might contribute differently when building an integrated representation. The general preference $\mathbf{O}_u$ considers the influence of the long-term user interest, while the current preference representation $\mathbf{C}_u$ exploits to utilize the basic item information of the current session.  Therefore, we design the following gating mechanism to form the final session preference representation $\mathbf{S}_{u}$: 
\begin{equation}
\begin{aligned}
    \alpha_{c}=\sigmoid(\mathbf{W}_s[\mathbf{C}_u || \mathbf{O}_u])\,,
\end{aligned}\label{eqn:graph}
\end{equation}
\begin{equation}
\begin{aligned}
    \mathbf{S}_{u}=\alpha_{c}\cdot \mathbf{C}_u + (1 - \alpha_{c})\cdot \mathbf{O}_u\,,
\end{aligned}\label{eqn:graph}
\end{equation}
where $\sigmoid(\cdot)$ is sigmoid function and $\mathbf{W}_{s} \in \mathbb{R}^{d \times 2d}$.

\subsection{Prediction and Training}\label{sec:search}
% To achieve personalized recommendation and improve the model performance, we combine the current session and the historical preference representation to generate the final user representation:
% \begin{equation}
% \begin{aligned}
%     \mathbf{S}=\mathbf{o}_{u}^s+\mathbf{o}_{u}^h\,.
% \end{aligned}\label{eqn:graph}
% \end{equation}

Based on the session preference  representation $\mathbf{S}_u$ and the initial embeddings of candidate items, we can compute the recommendation probability $\hat{\mathbf{y}}$ of candidate items in the current session:

\begin{equation}
\begin{aligned}
    \hat{{y}}_i=\text{softmax}(\mathbf{S}^T\mathbf{p}_{v_i}^{(0)})\,,
\end{aligned}\label{eqn:graph}
\end{equation}
where $\hat{{y}}_i \in \hat{\mathbf{y}}$ denotes the probability that the user will click on item $v_i \in V$ in the current session.

The objective function can be formulated as a cross entropy loss:% follows
\begin{equation}
\begin{aligned}                                                        
    \mathcal{L}(\hat{\mathbf{y}})=-\sum_{i=1}^{|V|}{y}_i\text{log}(\hat{{y}}_i)+(1-{y}_i)\text{log}(1-\hat{{y}_i})\,,
\end{aligned}\label{eqn:graph}
\end{equation}
where $\mathbf{y} \in \mathbb{R}^{|V|}$ is a one-hot vector of ground truth.

\section{Experiments}
To answer the following research questions, we conduct experiments on session-based recommendation to evaluate the performance of our method compared with other state-of-the-art models:
% \footnote{Our code and data will be released for research purpose.}
\begin{itemize}
    \item \textbf{RQ1:} How does our model perform compared with  state-of-the-art session-based recommender methods? 
    \item \textbf{RQ2:} How do different settings of HG-GNN and the heterogeneous graph construction influence the performance?
    \item \textbf{RQ3:} How do the hyper-parameters affect the effectiveness of our model?
    \item \textbf{RQ4:} Can our model make a personalized session-based recommendation? 
\end{itemize} 
\begin{table}[htbp]
    \caption{Statistics of datasets used in experiments.}
    \label{tab:dataset}
    \centering
    % \small
    \begin{tabular}{lrrr}
    \toprule
    Statistic& Last.fm & Xing & Reddit \\
    \midrule
    No. of users  & 992 & 11,479 & 18,271\\
    No. of items  & 38,615 &  59,121 & 27,452 \\
    No. of sessions  & 385,135 & 91,683 &1,135,488 \\
    Avg. of session length & 8.16 & 5.78 & 3.02\\
    Session per user & 373.19 & 7.99  & 62.15 \\
    No. of train sessions & 292,703 & 69,135  & 901,161 \\
    No. of test sessions & 92,432  & 22,548  & 234,327 \\
    
    \bottomrule
    \end{tabular}\vspace{-0.3cm}
\end{table}

\subsection{Experimental Setup}
\subsubsection{Dataset.} We conduct extensive experiments on three real-world datasets: \emph{Last.fm}, \emph{Xing} and \emph{Reddit} are widely used in the session-based recommendation research \citep{lessr,gce-gnn,guo2019streaming,ren2019repeatnet,H-RNN,A-PGNN}. These datasets both contain the basic user information that can support our work on personalized session-based recommendation.

\begin{itemize}
% \begin{inparaitem}
    \item \emph{Last.fm}\footnote{http://ocelma.net/MusicRecommendationDataset/lastfm-1K.html} contains the whole listening habits for nearly 1,000 users collected from Last.fm. In this work, we focus on the music artist recommendation. We keep the top 40,000 most popular artists and group interaction records in 8 hours from the same user as a session, following \citep{lessr,guo2019streaming}. 
    
    \item \emph{Xing}\footnote{http://2016.recsyschallenge.com/} collects the job postings from a social network platform, and contains interactions on job postings for 770,000 users. We split each user’s records into sessions manually by using the same approach as mentioned in \citep{H-RNN}.

    \item \emph{Reddit}\footnote{https://www.kaggle.com/colemaclean/subreddit-interactions}  contains tuples of user name, a subreddit where the user comment to a thread, and a interaction timestamp. We partitioned the interaction data into sessions by using a 60-minute time threshold following \citep{ludewig2019performance}.
    
\end{itemize}

To filter poorly informative sessions, we filtered out sessions having less than 3 interactions and kept users having 5 sessions or more to have sufficient historical sessions following \citep{H-RNN, A-PGNN}. We take last 20\% sessions as the test set, while the remaining sessions form the training set. Additionally, we filtered out interactions from the test set where the items do not appear in training set.

The statistics of mentioned datasets are summarized in \autoref{tab:dataset}. Referring to \citep{H-RNN, A-PGNN},  we segment the sessions of each user $S_u$ into historical sequences and item labels. For example, for the session data of user $u$ $S_u = \{\{v_{1,1},v_{1,2}\}, \{v_{2,1},v_{2,2}, v_{2,3}\}\}$, the historical sessions, current sessions and target label are set as $S_h^u= \{\{v_{1,1},v_{1,2}\}\}$, $S_c^u= \{{v_{2,1},v_{2,2}}\}$ and $label = v_{2,3}$. The target label is the next interacted item within the current session.

\subsubsection{Baseline Models.}
To evaluate the performance of our method, we compare it with several representative competitors, including the advanced GNN-based models and several personalized methods.

\begin{itemize}
% \begin{inparaitem}
    \item \textbf{Item-KNN} \citep{sarwar2001item} is a conventional item-to-item model which recommends items similar to the items in the session.
    % \footnote{https://github.com/hidasib/GRU4Rec}
    \item \textbf{GRU4Rec} \citep{gru4rec} employs the GRU to capture the representation of the item sequence through a session-parallel mini-batch training process.
    % \footnote{https://github.com/lijingsdu/sessionRec\_NARM}
    \item \textbf{NARM} \citep{narm} is also a RNN-based model which utilizes attention mechanism with GRU to generate session embedding.
    % \footnote{https://github.com/CRIPAC-DIG/SR-GNN}
    \item \textbf{SR-GNN} \citep{srgnn} converts session sequences into directed unweighted graphs and utilizes a  GGNN layer \citep{li2015gated} to learn the patterns of item transitions.
% \footnote{https://github.com/twchen/lessr}
    \item \textbf{LESSR} \citep{lessr} adds shortcut connections between items in the session and considers the sequence information in graph convolution by using GRU.
    % \footnote{https://github.com/CCIIPLab/GCE-GNN}
    \item \textbf{GCE-GNN} \citep{gce-gnn} aggregates the global context  and the item sequence in the current session to generate the session embedding through different level graph neural networks.
    % \footnote{https://github.com/mquad/hgru4rec}
    \item \textbf{H-RNN} \citep{H-RNN} is a RNN-based personalized method which utilizes a hierarchical RNNs consist of a session-based and a user-level RNN to model the cross-session user interests. 
    % \footnote{https://github.com/CRIPAC-DIG/A-PGNN}
    \item \textbf{A-PGNN} \citep{A-PGNN} converts all sessions of each user into a graph and employ the GGNN model to learn the item transitions. 
    Besides, As a personalized recommender, A-PGNN utilizes the attention mechanism to explicitly model the effect of the user’s historical interests on the current session. 
    
\end{itemize}

\subsubsection{Evaluation Metrics.}
To evaluate the recommendation performance, we employ two widely used metrics: Hit Ratio (HR@$k$)  and  Mean Reciprocal Rank (MRR@$k$) following  \citep{srgnn,gce-gnn}, where $k =\{5, 10\}$. The average results over all test users are reported.

\subsubsection{Implementation Details.}
We implement the proposed model based on Pytorch and DGL. We employ the Adam optimizer with the initial learning rate $\eta$.%which will decay by 0.1 after every 3 epochs following \citep{srgnn,gce-gnn}
 The mini-batch size is set to be 512 for all models. We employ the grid search to find the optimal hyper-parameters by taking 10\% of training data as the validation set. The ranges of the hyper-parameters are: $\{32, 64, 96, 128, 256\}$ for embedding size $d$,  $\{4,8,12,16\}$ for sampling size $S$ and $\{5, 10, ..., 30\}$ for the number of similarity items $K$. For the GNN-based models, we search the total number of GNN layers in $\{1,2,3,4\}$. We report the result of each model under its optimal hyper-parameter settings. 
%  For the global graph construction, we set the sampling size $S$ and the number of similarity items $K$ as hyper-parameters.  
% Furthermore, we adopt an early stopping strategy, i.e. stopping the training process if metric HR@10 does not increase for 5 successive epochs. 

% \end{inparaitem}
\subsection{Model Comparison (RQ1)}

\begin{table*}[t]
    \centering
    \caption{Experimental results (\%) of different models in terms of HR@\{5, 10\}, and MRR@\{5, 10\} on three datasets. The * means the best results on baseline methods.  $Improv.$ means improvement over the state-of-art methods. $^\dagger$  denotes a significant improvement of our model over the best baseline using a paired t-test ($p$ < 0.01).}
    \label{tab:overall}
    \begin{tabular}{p{1.4cm}<{\centering}p{0.9cm}<{\centering}p{0.9cm}<{\centering}p{0.9cm}<{\centering}p{0.9cm}<{\centering}p{0.05cm}p{0.9cm}<{\centering}p{0.9cm}<{\centering}p{0.9cm}<{\centering}p{0.9cm}<{\centering}p{0.05cm}p{0.9cm}<{\centering}p{0.9cm}<{\centering}p{0.9cm}<{\centering}p{0.9cm}<{\centering}}%{lrrrrrrrrrrrr}
    \toprule
    \multirow{2}{*}{ \bfseries Models}& \multicolumn{4}{c}{ \bfseries Last.fm }& & \multicolumn{4}{c}{\bfseries Xing}& &\multicolumn{4}{c}{\bfseries Reddit} \\
    \cline{2-5}
    
    \cline{7-10}
    
    \cline{12-15}
    
    &HR@5&HR@10&MRR@5&MRR@10&&HR@5&HR@10&MRR@5&MRR@10&&HR@5&HR@10&MRR@5&MRR@10\\
   % \cline{1-7}
    \midrule
    ItemKNN& 6.73 & 10.90 & 4.02 & 4.81 & & 8.79 & 11.85 & 5.01 &  5.42 & & 21.71 & 30.32 & 11.74 & 12.88\\

    %\cline{1-7}
    GRU4Rec&8.47&12.86&4.71&5.29&  & 10.35 & 13.15 & 5.94 & 6.36 & & 33.72 & 41.73 & 24.36 & 25.42\\
    
    %\cline{1-7}
    NARM&10.29&15.03&6.09&6.71& &13.51 & 17.31 & 8.87 & 9.37 & & 33.25 & 40.52 & 24.56 & 25.52  \\

    %\cline{1-7}
    SR-GNN& 11.89 & 16.90 & 7.23 & 7.85  & & 13.38 & 16.71  & 8.95 & 9.39 & & 34.96 & 42.38 & 25.90 & 26.88\\
    
    %\cline{1-7}
    LESSR&12.96* &17.88&8.24* & 8.82*& & 14.84 & 16.77 & 11.98* & 12.13* & & 36.03 & 43.27 & 26.45 & 27.41 \\
    
    GCE-GNN&12.83 & 18.28* & 7.60 & 8.32 & & 16.98* & 20.86* & 11.14 & 11.65 & & 36.30 & 45.16 & 26.65 & 27.70 \\
    \midrule
    
    H-RNN &10.92 & 15.83 & 6.71 & 7.39 & &10.72 & 14.36 &7.22 & 7.74& & 44.76  & 53.44 &32.13 & 33.29\\
    
    A-PGNN & 12.10 & 17.13 & 7.37 & 8.01&  & 14.23 & 17.01 & 10.26 & 10.58 & & 49.10* & 58.23* & 33.54* & 34.62*\\

    %\cline{1-7}
    \midrule

    HG-GNN&{\bfseries 13.09$^\dagger$} &{\bfseries 19.39$^\dagger$}& { 7.35} &{  8.18}& &{\bfseries 17.25$^\dagger$}&{ 20.30} &{\bfseries 12.23$^\dagger$}&{\bfseries 12.79$^\dagger$} & &{\bfseries 51.08$^\dagger$}&{\bfseries 60.51$^\dagger$} &{\bfseries 35.46$^\dagger$}&{\bfseries 36.89$^\dagger$}\\

    {$Improv.$}& 1.00\% & 6.07\% & - & - & & 1.59\% & - & 2.09\% & 5.44\% & & 4.03\% & 3.92\% & 5.72\% & 6.56\%    \\
    \bottomrule
    \end{tabular}\vspace{-0.3cm}
\end{table*}

To demonstrate the overall performance of the proposed model, we compared it with the state-of-the-art recommendation methods. We can obtain the following important observations from the comparison results shown in \autoref{tab:overall}.

First, HG-GNN outperforms other SOTA models on all datasets on most metrics, demonstrating the superiority of our model. Among the comparison methods, the performance of the conventional method ( Item-KNN ) is not competitive. Meanwhile, all deep learning-based methods achieve better performance than Item-KNN. 
% It proves that deep learning can capture the complex patterns of item transitions in the session, and is efficient for session-based recommendation.

Besides, we compare our model with the advanced GNN-based methods. These methods generally achieve better performance in contrast with the RNN-based methods. SR-GNN converts the session sequence to a directed graph and utilizes the GNN to encode the session graph. As the variant methods of SR-GNN, LESSR also achieves promising results. It demonstrates that GNN model has a stronger capability of modeling session sequences than the RNN model in session-based recommendation tasks. Furthermore, GCE-GNN not only utilizes the session graph for the item sequence, but also constructs an undirected global graph from all sessions following interaction sequences. GCE-GNN outperforms SR-GNN on both datasets, showing the importance of global context information for the session-based recommendation.

Compared with the state-of-the-art personalized session-based approaches i.e. H-RNN and A-PGNN, our approach achieves a significant performance improvement on all metrics consistently. Specifically, HG-GNN outperforms A-PGNN by 4.03\% in terms of HR@5 and 5.72\% in terms of MRR@5 on \emph{Reddit}. We can attribute the success of A-PGNN to the ability to model explicitly the effect of the user’s historical interests on the current session. However,  these  methods only focus on the cross-session information of the current user while ignore the influence of global context information of historical sessions. In contrast, HG-GNN overcomes this deficiency by forming the heterogeneous global session graph. Through the global graph, we effectively organize historical sessions. The design of heterogeneous global graph and effective preference modeling together contribute to the remarkable performance.
%  we learn the user and item embeddings with rich semantics through the HGNN layer.

\begin{table}[htbp]
    \centering
    \caption{Runtime (seconds) of each training epoch.}
    \label{tab:runtime}
    \begin{tabular}{lrrr}
    \toprule
    {\bfseries Method} & {\bfseries Last.fm} &{\bfseries Xing} &{\bfseries Reddit} \\
   % \cline{1-7}
    \midrule
    A-PGNN & 995.53 & 118.18 & 1,054.20\\
    % LESSR & 612.80 & 76.68 & 790.83 \\
    GCE-GNN & 521.23 & 76.35 & 605.37\\
    \midrule
    HG-GNN & 458.56 & 61.23 & 576.13\\
    \bottomrule
    \end{tabular}
\end{table}

We also record the runtime of some GNN-based methods and the proposed HG-GNN approach. We implement models with the same 128-dimensional embedding vectors and the same batch size, and test them on the same GPU server (Tesla V100 DGXS 32GB) with sufficient resources. We only record the training time of each epoch, and do not consider the process of data loading and testing such as the cost of graph creation. In other words, we measure the runtime of the model network operations. The average running time of 10 epochs on three datasets are reported in \autoref{tab:runtime}. 
It illustrates that HG-GNN is more efficient than A-PGNN and GCE-GNN. We argue that this is because the A-PGNN contains the extra transformer module which conducts expensive computation, compared with HG-GNN. In addition to the GNN layer on the global graph, GCE-GNN also involves extra GNN layers on the session graph which leading to higher time consumption than HG-GNN.

\subsection{Ablation and Effectiveness Analyses (RQ2)}\label{sec:ablation}

In this subsection, we conduct some ablation studies on the proposed model to investigate the effectiveness of some designs.

\subsubsection{Impact of different Layers}

In this part, we compare our method with different variants to verify the effectiveness of the critical components of HG-GNN. Specifically, we remove critical modules of HG-GNN to observe changes about model performance. The results of ablations studies are shown in the \autoref{tab:layer}. We observe that the HGNN module is pivotal for the model performance by seeing "w/o HGNN module".  For the part of Personalized Session Encoder, the removal of current preference (CPL) and the general preference learning (GPL) module both lead to poor performance. Through the above comparison and analysis, we can conclude that the main components design of HG-GNN are effective.

% \autoref{tab:layer} shows the different impact of the layers of our model. The experiment of  without PSE layer shows that we only utilize the refined item embeddings from HGNN layer to generate the sequence representation. Both the HGNN layer and the PSE layer contribute to model performance, with the former being more critical to recommendation performance.

\begin{table*}[t]
    \caption{Impact of different layer of HG-GNN. }
    \label{tab:layer}
    \centering
    % \small
    %\setlength{\tabcolsep}{2pt}
    \begin{tabular}{lccccrcccc}
    \toprule
   \multirow{2}{*}{ \bfseries Model setting}& \multicolumn{4}{c}{\bfseries Last.fm }& &\multicolumn{4}{c}{\bfseries Xing }\\
     \cline{2-5}
    \cline{7-10}
     &HR@5&HR@10&MRR@5&MRR@10& &HR@5&HR@10&MRR@5&MRR@10\\
    \midrule
    w/o HGNN module & \makecell[c]{11.98} & \makecell[c]{17.42} & \makecell[c]{6.81} & \makecell[c]{7.53} & & \makecell[c]{16.54} & \makecell[c]{20.14} & \makecell[c]{11.76} & \makecell[c]{12.24}  \\
    % \cline{2-10}
   
    w/o CPL module & \makecell[c]{12.83} & \makecell[c]{19.16}&\makecell[c]{7.07} & \makecell[c]{7.91}& &\makecell[c]{13.16} & \makecell[c]{16.53} & \makecell[c]{8.70} & \makecell[c]{9.15}  \\
    % \cline{2-10}
    % w/o user embedding &\makecell[c]{12.48\\(-5.10\%)}&\makecell[c]{18.06\\(-7.15\%)}& \makecell[c]{7.12\\(-4.43\%)}& \makecell[c]{7.86\\(-5.07\%)}&  &\makecell[c]{16.72\\(-3.85\%)}&\makecell[c]{20.24\\(-2.50\%)}&\makecell[c]{12.00\\(-5.21\%)}&\makecell[c]{12.23\\(-6.71\%)} \\
    %  \cline{2-10}
    w/o GPL module  &\makecell[c]{12.96}&\makecell[c]{19.22}&\makecell[c]{7.13}&\makecell[c]{8.04}& & \makecell[c]{17.24} & \makecell[c]{20.38} & \makecell[c]{12.21} & \makecell[c]{12.74}  \\
    \midrule
    HG-GNN&13.09& 19.39 & 7.35 & 8.18 & & 17.25 & 20.30 & 12.23 & 12.79\\
    \bottomrule
    \end{tabular}\vspace{-0.3cm}
\end{table*}

\subsubsection{Impact of different Global Graph Design.}
We next conduct experiments to evaluate the effectiveness of the proposed heterogeneous global graph. Specifically, we remove different types of edges and nodes in the graph to observe  their impact on the model. The experimental results are shown in \autoref{tab:graph_relation}.

We can see that the deletion nodes or edges can bring performance loss. Removing the user nodes leads to  the most significant performance drop, demonstrating the importance of user information and user-item interaction for recommendation. It is worth noting that our method without user nodes is still competitive, compared with other methods in \autoref{tab:overall}. It also can be observed that without the similar edges constructed by co-occurrence, the performance  declines to a certain extent. This indicates that constructing extra edges through the global co-occurrence information between is effective and useful for session-based recommendations. Additionally, removing the $in$ or $out$ edges can bring more performance loss than removing the $similar$ edges. It demonstrates that 
adjacent interactions relationship and the global co-occurrence relationship complement each other, and the adjacent interactions relationship is more important in the session-based recommendation task.

\begin{table}[htbp]
    \caption{Comparison w.r.t. different graph design on \emph{Xing}. }
    \label{tab:graph_relation}
    \centering
    % \small
    \setlength{\tabcolsep}{2pt}
    \begin{tabular}{lrrrr}
    \toprule
   
    {\bfseries Graph Setting}& HR@5&HR@10&MRR@5&MRR@10\\
    \midrule
    {w/o \emph{user} nodes}&17.03&19.82&12.05&12.86\\
    
    {w/o \emph{in}  edges} &17.14 & 20.02 & 12.12 & 12.93 \\
    
    {w/o \emph{out}  edges} &17.11& 20.00 & 12.02 & 12.87 \\
    
    {w/o \emph{similar} edges}&17.24&20.09&12.32&13.04\\
  
    \midrule
     HG-GNN&17.25& 20.30 & 12.23 & 12.79 \\
    
    \bottomrule
    \end{tabular}
\end{table}

% \begin{figure*}[t]
%     \centering
%     \begin{subfigure}{0.1\linewidth}
%         \includegraphics[width=\textwidth]{img/gnn_layer_lastfm.pdf}
%     \end{subfigure}
%     \begin{subfigure}{0.1\linewidth}
%         \includegraphics[width=\textwidth]{img/embedding_size_lastfm.pdf}
%     \end{subfigure}
%     \begin{subfigure}{0.1\linewidth}
%         \includegraphics[width=\textwidth]{img/rank_gnn_128.pdf}
%     \end{subfigure}
%     \begin{subfigure}{0.1\linewidth}
%         \includegraphics[width=\textwidth]{img/embedding_size_lastfm.pdf}
%     \end{subfigure}

%     \caption{Comparisons of search efficiency. Left: validation (top) and test (bottom) error of architectures searched on NAS-Bench-101; right: validation (top) and test (bottom) error of architectures searched on NAS-Bench-201.}
%     \label{fig:search}
% \end{figure*}

\subsection{Hyper-parameters Study (RQ3) }\label{sec:hyper-res}
% We evaluate the efficiency of our method on \emph{NAS-Bench-101} and \emph{NAS-Bench-201} benchmarks.
In this subsection, we perform experiments to explore how the hyper-parameters like sampling size $S$  influence the model performance. Due to space limitation, we only show parts of the hyper-parameter experimental results.

\subsubsection{Effect of the sampling size.}
\begin{figure}[t]
    \centering
    \begin{subfigure}{0.45\linewidth}
        \includegraphics[width=\textwidth]{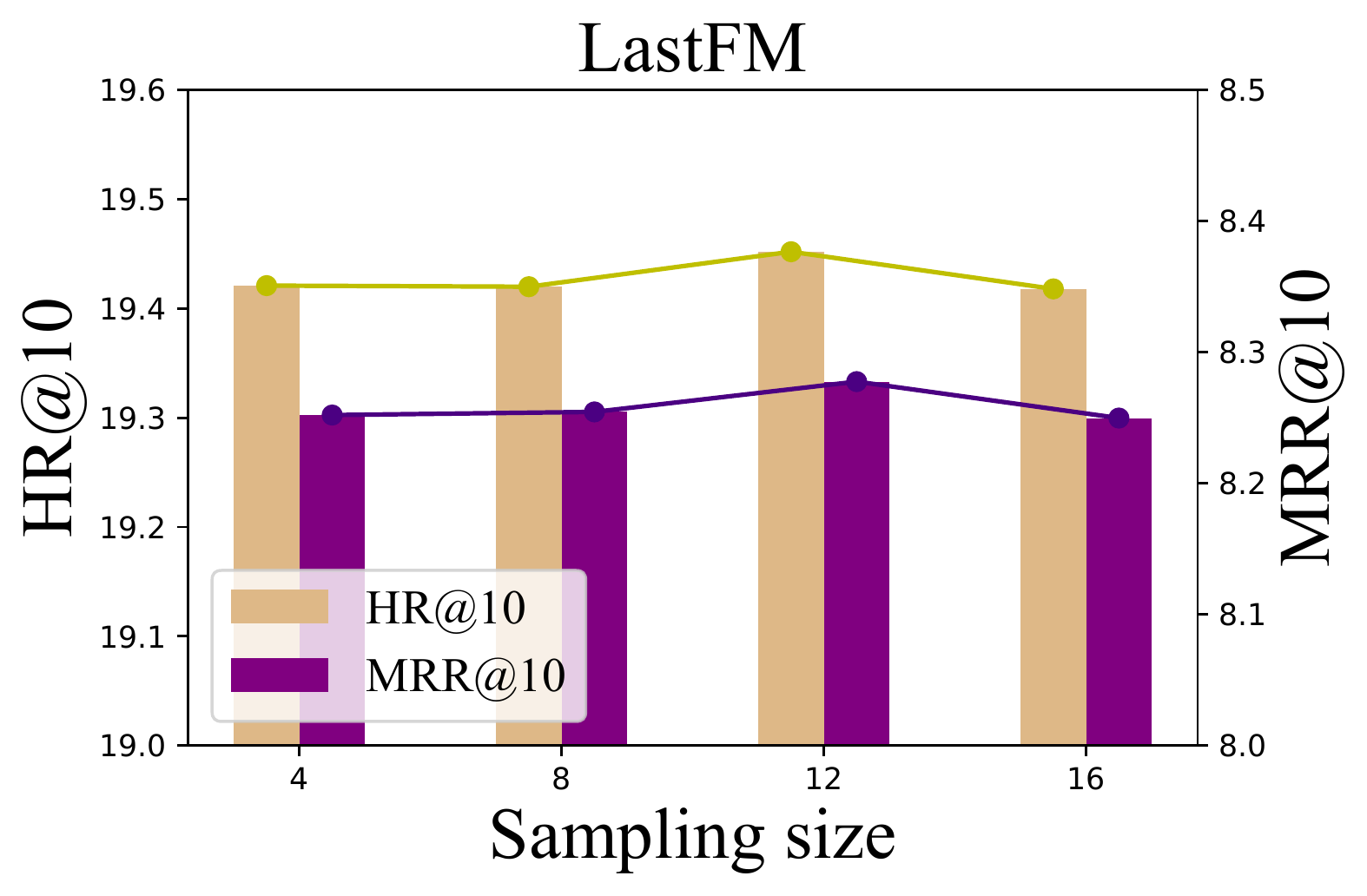}
    \end{subfigure}
    \begin{subfigure}{0.45\linewidth}
        \includegraphics[width=\textwidth]{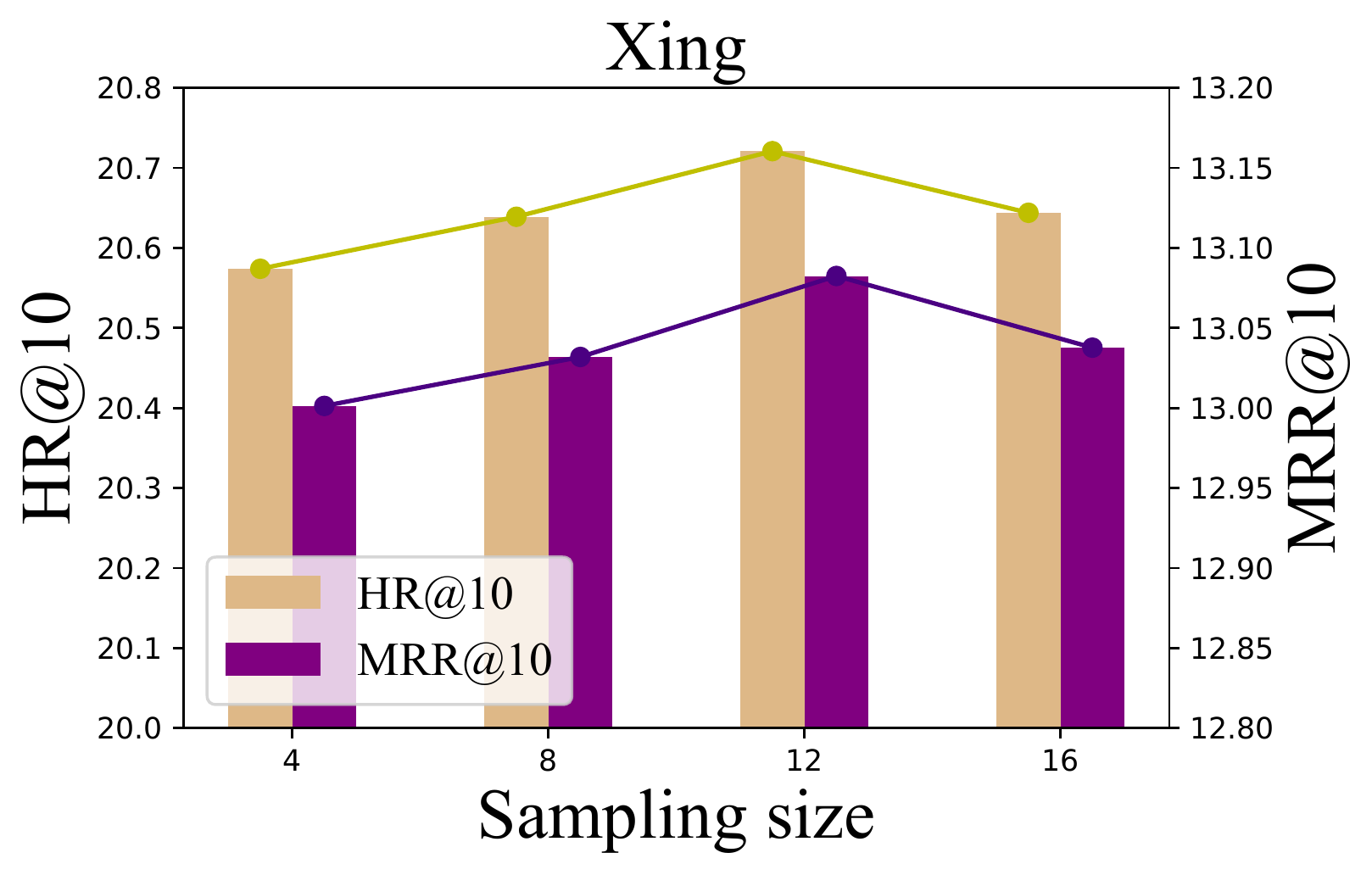}
    \end{subfigure}

    \caption{Comparison w.r.t. sampling size $S$.}
    \label{fig:hyper-sz}
\end{figure}
% Performance comparison w.r.t. sampling size $S$ in the global graph construction.
In the process of building the global graph, due to calculation efficiency and information validity,  we sample $S$ edges for each item node according to the weight of the adjacent edges $r_{in}$ or $r_{out}$. Thus, $S$ is the pivotal hyper-parameter for our model. The experimental results are shown in the \autoref{fig:hyper-sz}. As can be seen that the model performance reaches the highest value when $S$ is 8 for the \emph{Last.fm} and \emph{Xing} dataset. The metrics change in the figure can show that too big or too small sampling size will bring loss to the model effect. The most suitable sampling size can achieve the balance between effective information and irrelevant information, so as to achieve the best performance.

\subsubsection{Effect of the Top-$K$ similar items.}

\begin{figure}[t]
    \centering
    \begin{subfigure}{0.45\linewidth}
        \includegraphics[width=\textwidth]{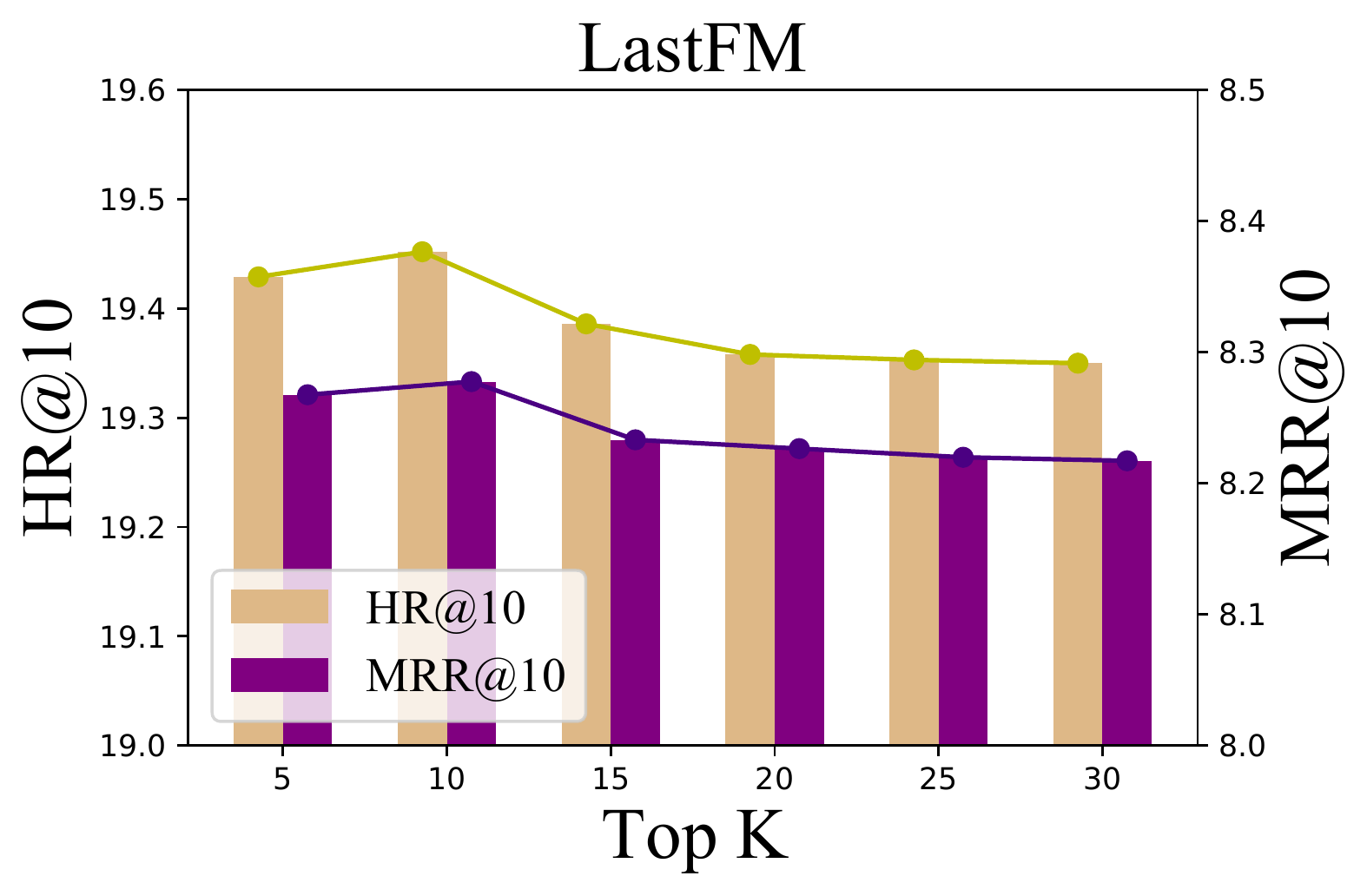}
    \end{subfigure}
    \begin{subfigure}{0.45\linewidth}
        \includegraphics[width=\textwidth]{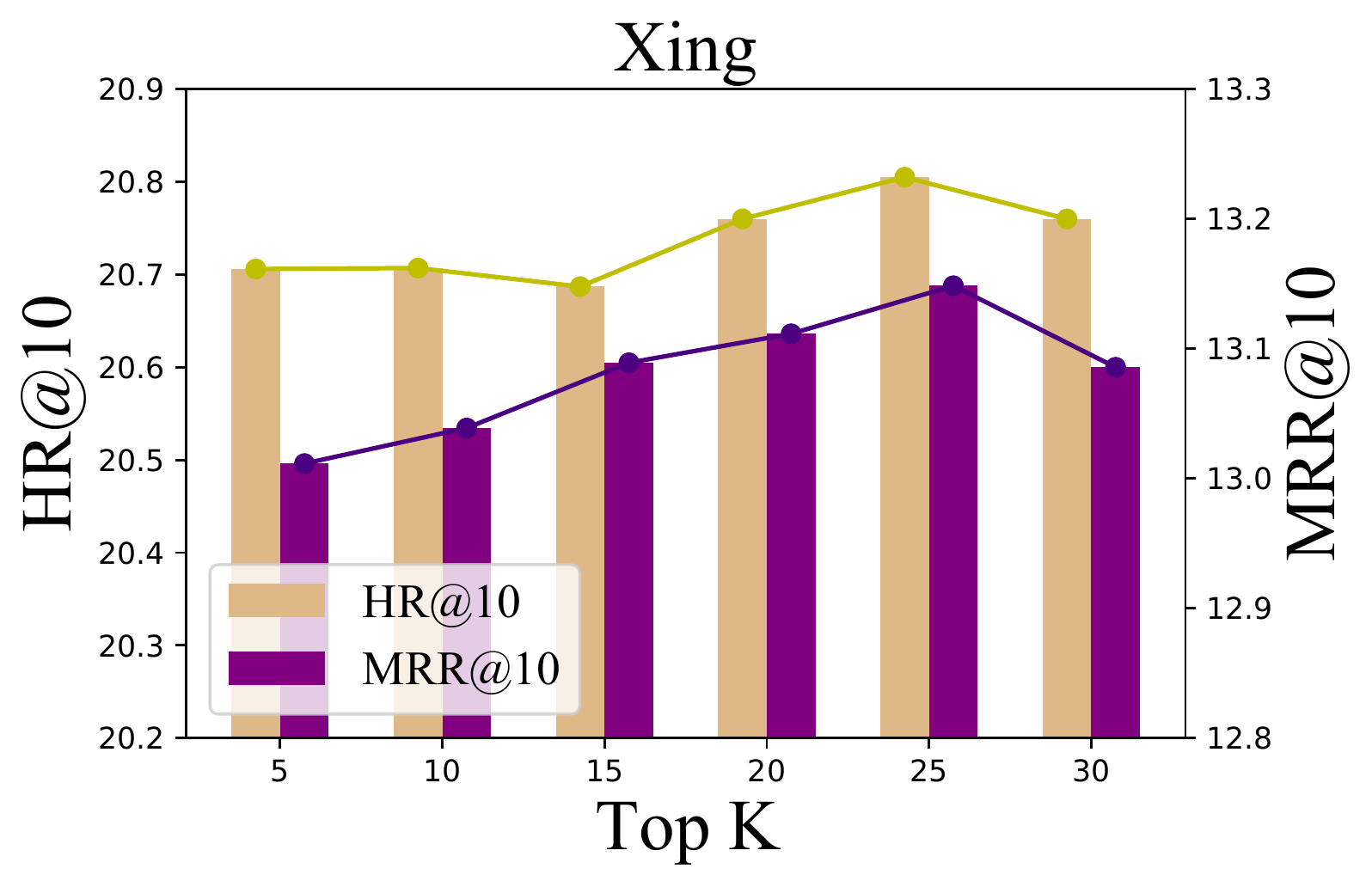}
    \end{subfigure}

    \caption{Comparison w.r.t. the Top-$K$ similar items.}
    \label{fig:hyper-topk}\vspace{-0.3cm}
\end{figure}
% Performance comparison w.r.t. the Top-$K$ similar items in the global graph construction.

% In addition to the sampling size discussed above, 
The number of similar items is another key hyper-parameter for the global graph. For each item, we select the top-$K$ similar items based on the global co-occurrence information to construct the similar edges $r_{similar}$. In practice, the number of $r_{similar}$ edges for each item node is cut off based on the size of adjacent interaction items. The experimental results are shown in the \autoref{fig:hyper-topk}. It is clear that the optimal $K$ value is quite different on different datasets. Since the relationship of similar items depends on the global interaction, the distribution of similar relationships on different datasets will vary greatly. Therefore, this hyper-parameter is relatively sensitive to data.

% \begin{figure}[t]
%     \centering
  
%     \begin{subfigure}{0.45\linewidth}
%         \includegraphics[width=\textwidth]{img/embedding_size_lastfm.pdf}
%     \end{subfigure}
%     \begin{subfigure}{0.45\linewidth}
%         \includegraphics[width=\textwidth]{img/embedding_size_nowplaying.pdf}
%     \end{subfigure}
%     \caption{Performance comparison w.r.t. different embedding size.}
%     \label{fig:hyper-param2}
% \end{figure}

\begin{figure}[t]
    \centering
    \includegraphics[width=0.45\textwidth]{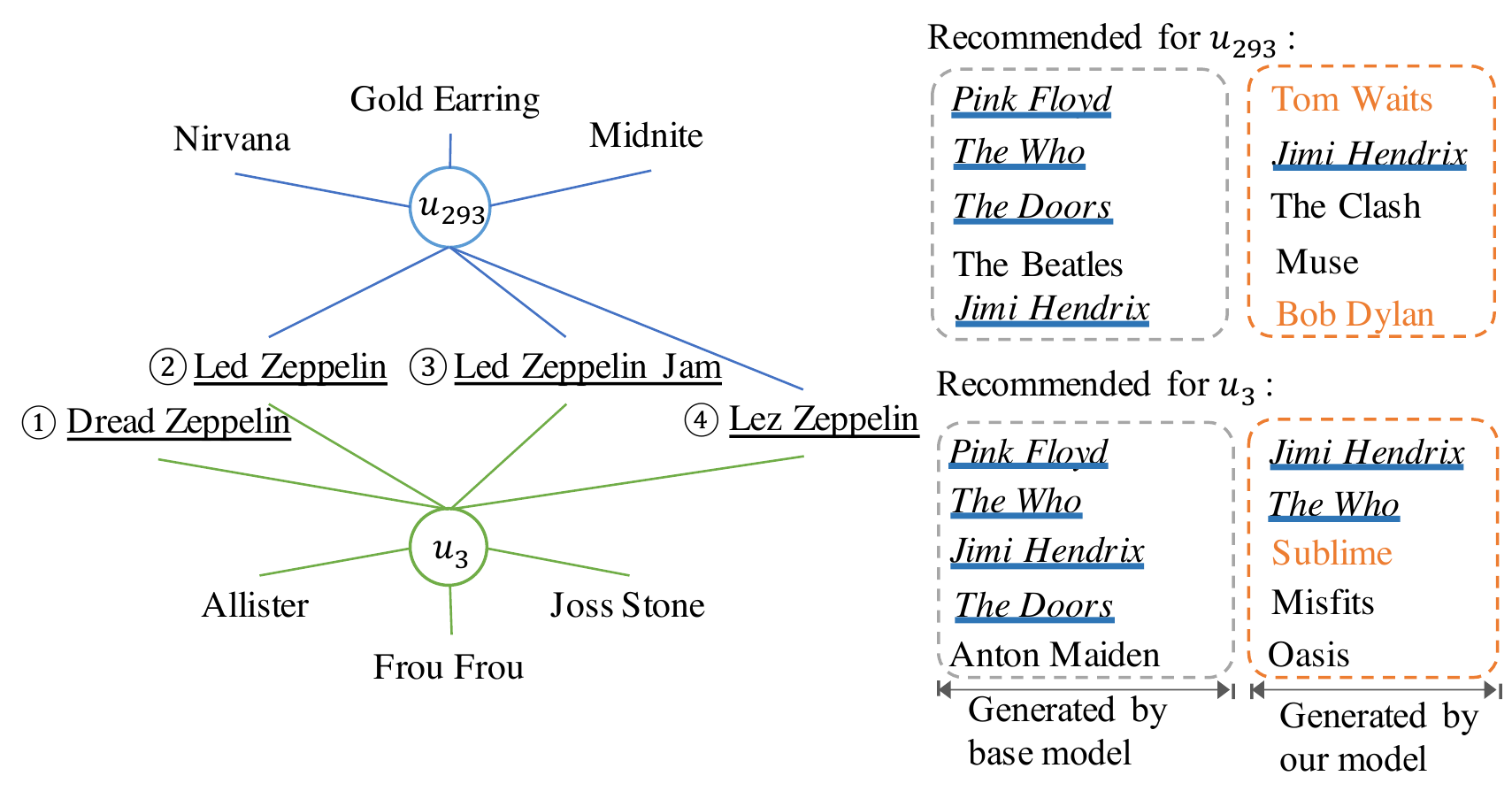}
    \caption{A case study of two users from \emph{Last.fm} data for music artist recommendation.  Artists underlined in blue denote the same recommendation results of two user, and artists in yellow indicate personalized results.}
    \label{fig:case_study)}\vspace{-0.7cm}
\end{figure}
% Two users have different music preferences but similar ongoing session sequences underlined in black. This figure presents the difference between the recommendation results (top 5 artists) generated by our model and the base model.

\subsection{Case Study (RQ4) }\label{sec:case_study}
To illustrate the personalized result of our model intuitively, we present a music artist recommendation case in \autoref{fig:case_study)}. We select two users from \emph{Last.fm}, who have different listening history. Given their similar current session sequences (sessions both contain the same artists underlined in black), our model generates different recommendation lists than non-personalized base model (SR-GNN).

For user $u_{293}$ and $u_3$, they both have different music preferences but have partially overlapping listening histories. The former prefers classic music and rock music e.g. \emph{Gold Earring}, while the latter is younger and loves pop and rock music, such as \emph{Joss Stone}. The current sessions share  \emph{Led Zeppelin} (a famous rock band). For such similar sessions, artists recommended by base model overlap much for two users (here we only take the top 5 recommendation results). Our model takes into account user preferences, so it can recommend musicians who are more relevant to the current session such as \emph{Jimi Hendrix}, as well as artists that meet other preferences of users, such as  \emph{Sublime} (a ska punk band) for $u_3$. In summary, our model can achieve more accurate and personalized recommendations.

\section{Conclusion}
In this paper, we proposed a heterogeneous global graph neural network for personalized session-based recommendation. Contrasting with previous methods, we considered the impact of historical interactions of users and build a heterogeneous global graph that consists of historical user-item interactions, item transitions and global co-occurrence information. Furthermore, we proposed a graph augmented hybrid encoder which consists of a heterogeneous graph neural network and personalized session encoder to capture the user preference representation comprehensively. In the experiments, our model outperforms other state-of-the-art session-based models, showing the effectiveness of our model. 
% For future work, we are interested in applying HG-GNN to other tasks like social recommendation \citep{chen2021efficient} to investigate its scalability.

\begin{acks}
The work is partially supported by the National Nature Science Foundation of China (No. 61976160, 61976158, 61906137), Shanghai Science and Technology Plan Project (No. 21DZ1204800) and Technology research plan project of Ministry of Public and Security (Grant No. 2020JSYJD01).
\end{acks}
%\clearpage
\bibliographystyle{ACM-Reference-Format}
\balance
\bibliography{ref}


\begin{thebibliography}{35}

%%% the Web address.

\ifx \showCODEN    \undefined \def \showCODEN     #1{\unskip}     \fi
\ifx \showDOI      \undefined \def \showDOI       #1{#1}\fi
\ifx \showISBNx    \undefined \def \showISBNx     #1{\unskip}     \fi
\ifx \showISBNxiii \undefined \def \showISBNxiii  #1{\unskip}     \fi
\ifx \showISSN     \undefined \def \showISSN      #1{\unskip}     \fi
\ifx \showLCCN     \undefined \def \showLCCN      #1{\unskip}     \fi
\ifx \shownote     \undefined \def \shownote      #1{#1}          \fi
\ifx \showarticletitle \undefined \def \showarticletitle #1{#1}   \fi
\ifx \showURL      \undefined \def \showURL       {\relax}        \fi
% invisible to TeX
\providecommand\bibfield[2]{#2}
\providecommand\bibinfo[2]{#2}
\providecommand\natexlab[1]{#1}
\providecommand\showeprint[2][]{arXiv:#2}

\bibitem[\protect\citeauthoryear{Chen and Wong}{Chen and Wong}{2020}]%
        {lessr}
\bibfield{author}{\bibinfo{person}{Tianwen Chen} {and} \bibinfo{person}{Raymond
  Chi-Wing Wong}.} \bibinfo{year}{2020}\natexlab{}.
\newblock \showarticletitle{Handling Information Loss of Graph Neural Networks
  for Session-based Recommendation}. In \bibinfo{booktitle}{\emph{KDD}}.
  \bibinfo{pages}{1172--1180}.
\newblock


\bibitem[\protect\citeauthoryear{Chen and Wong}{Chen and Wong}{2021}]%
        {chen2021efficient}
\bibfield{author}{\bibinfo{person}{Tianwen Chen} {and} \bibinfo{person}{Raymond
  Chi-Wing Wong}.} \bibinfo{year}{2021}\natexlab{}.
\newblock \showarticletitle{An Efficient and Effective Framework for
  Session-based Social Recommendation}. In \bibinfo{booktitle}{\emph{WSDM}}.
  \bibinfo{pages}{400--408}.
\newblock


\bibitem[\protect\citeauthoryear{Chen, Wu, and Zaki}{Chen
  et~al\mbox{.}}{2019}]%
        {chen2019reinforcement}
\bibfield{author}{\bibinfo{person}{Yu Chen}, \bibinfo{person}{Lingfei Wu},
  {and} \bibinfo{person}{Mohammed~J Zaki}.} \bibinfo{year}{2019}\natexlab{}.
\newblock \showarticletitle{Reinforcement learning based graph-to-sequence
  model for natural question generation}. In \bibinfo{booktitle}{\emph{ICLR
  2020}}.
\newblock


\bibitem[\protect\citeauthoryear{Chen, Wu, and Zaki}{Chen
  et~al\mbox{.}}{2020}]%
        {chen2020iterative}
\bibfield{author}{\bibinfo{person}{Yu Chen}, \bibinfo{person}{Lingfei Wu},
  {and} \bibinfo{person}{Mohammed~J Zaki}.} \bibinfo{year}{2020}\natexlab{}.
\newblock \showarticletitle{Iterative Deep Graph Learning for Graph Neural
  Networks: Better and Robust Node Embeddings}. In
  \bibinfo{booktitle}{\emph{NeurIPS 2020}}.
\newblock


\bibitem[\protect\citeauthoryear{Dias and Fonseca}{Dias and Fonseca}{2013}]%
        {dias2013improving}
\bibfield{author}{\bibinfo{person}{Ricardo Dias} {and}
  \bibinfo{person}{Manuel~J Fonseca}.} \bibinfo{year}{2013}\natexlab{}.
\newblock \showarticletitle{Improving music recommendation in session-based
  collaborative filtering by using temporal context}. In
  \bibinfo{booktitle}{\emph{2013 IEEE 25th international conference on tools
  with artificial intelligence}}. IEEE, \bibinfo{pages}{783--788}.
\newblock


\bibitem[\protect\citeauthoryear{Guo, Yin, Wang, Chen, Zhou, and Quoc
  Viet~Hung}{Guo et~al\mbox{.}}{2019}]%
        {guo2019streaming}
\bibfield{author}{\bibinfo{person}{Lei Guo}, \bibinfo{person}{Hongzhi Yin},
  \bibinfo{person}{Qinyong Wang}, \bibinfo{person}{Tong Chen},
  \bibinfo{person}{Alexander Zhou}, {and} \bibinfo{person}{Nguyen Quoc
  Viet~Hung}.} \bibinfo{year}{2019}\natexlab{}.
\newblock \showarticletitle{Streaming session-based recommendation}. In
  \bibinfo{booktitle}{\emph{KDD}}. \bibinfo{pages}{1569--1577}.
\newblock


\bibitem[\protect\citeauthoryear{Hamilton, Ying, and Leskovec}{Hamilton
  et~al\mbox{.}}{2017}]%
        {hamilton2017sage}
\bibfield{author}{\bibinfo{person}{William~L Hamilton}, \bibinfo{person}{Rex
  Ying}, {and} \bibinfo{person}{Jure Leskovec}.}
  \bibinfo{year}{2017}\natexlab{}.
\newblock \showarticletitle{Inductive representation learning on large graphs}.
\newblock \bibinfo{journal}{\emph{arXiv preprint arXiv:1706.02216}}
  (\bibinfo{year}{2017}).
\newblock


\bibitem[\protect\citeauthoryear{He, Deng, Wang, Li, Zhang, and Wang}{He
  et~al\mbox{.}}{2020}]%
        {he2020lightgcn}
\bibfield{author}{\bibinfo{person}{Xiangnan He}, \bibinfo{person}{Kuan Deng},
  \bibinfo{person}{Xiang Wang}, \bibinfo{person}{Yan Li},
  \bibinfo{person}{Yongdong Zhang}, {and} \bibinfo{person}{Meng Wang}.}
  \bibinfo{year}{2020}\natexlab{}.
\newblock \showarticletitle{Lightgcn: Simplifying and powering graph
  convolution network for recommendation}. In
  \bibinfo{booktitle}{\emph{SIGIR}}. \bibinfo{pages}{639--648}.
\newblock


\bibitem[\protect\citeauthoryear{Hidasi, Karatzoglou, Baltrunas, and
  Tikk}{Hidasi et~al\mbox{.}}{2015}]%
        {gru4rec}
\bibfield{author}{\bibinfo{person}{Bal{\'a}zs Hidasi},
  \bibinfo{person}{Alexandros Karatzoglou}, \bibinfo{person}{Linas Baltrunas},
  {and} \bibinfo{person}{Domonkos Tikk}.} \bibinfo{year}{2015}\natexlab{}.
\newblock \showarticletitle{Session-based recommendations with recurrent neural
  networks}.
\newblock \bibinfo{journal}{\emph{arXiv preprint arXiv:1511.06939}}
  (\bibinfo{year}{2015}).
\newblock


\bibitem[\protect\citeauthoryear{Kipf and Welling}{Kipf and Welling}{2016}]%
        {kipf2016semi}
\bibfield{author}{\bibinfo{person}{Thomas~N Kipf} {and} \bibinfo{person}{Max
  Welling}.} \bibinfo{year}{2016}\natexlab{}.
\newblock \showarticletitle{Semi-supervised classification with graph
  convolutional networks}.
\newblock \bibinfo{journal}{\emph{arXiv preprint arXiv:1609.02907}}
  (\bibinfo{year}{2016}).
\newblock


\bibitem[\protect\citeauthoryear{Latifi, Mauro, and Jannach}{Latifi
  et~al\mbox{.}}{2021}]%
        {latifi2021session}
\bibfield{author}{\bibinfo{person}{Sara Latifi}, \bibinfo{person}{Noemi Mauro},
  {and} \bibinfo{person}{Dietmar Jannach}.} \bibinfo{year}{2021}\natexlab{}.
\newblock \showarticletitle{Session-aware recommendation: A surprising quest
  for the state-of-the-art}.
\newblock \bibinfo{journal}{\emph{Information Sciences}}  \bibinfo{volume}{573}
  (\bibinfo{year}{2021}), \bibinfo{pages}{291--315}.
\newblock


\bibitem[\protect\citeauthoryear{Li, Ren, Chen, Ren, Lian, and Ma}{Li
  et~al\mbox{.}}{2017}]%
        {narm}
\bibfield{author}{\bibinfo{person}{Jing Li}, \bibinfo{person}{Pengjie Ren},
  \bibinfo{person}{Zhumin Chen}, \bibinfo{person}{Zhaochun Ren},
  \bibinfo{person}{Tao Lian}, {and} \bibinfo{person}{Jun Ma}.}
  \bibinfo{year}{2017}\natexlab{}.
\newblock \showarticletitle{Neural attentive session-based recommendation}. In
  \bibinfo{booktitle}{\emph{CIKM}}. \bibinfo{pages}{1419--1428}.
\newblock


\bibitem[\protect\citeauthoryear{Li, Tarlow, Brockschmidt, and Zemel}{Li
  et~al\mbox{.}}{2015}]%
        {li2015gated}
\bibfield{author}{\bibinfo{person}{Yujia Li}, \bibinfo{person}{Daniel Tarlow},
  \bibinfo{person}{Marc Brockschmidt}, {and} \bibinfo{person}{Richard Zemel}.}
  \bibinfo{year}{2015}\natexlab{}.
\newblock \showarticletitle{Gated graph sequence neural networks}.
\newblock \bibinfo{journal}{\emph{arXiv preprint arXiv:1511.05493}}
  (\bibinfo{year}{2015}).
\newblock


\bibitem[\protect\citeauthoryear{Liang, Li, Li, Gu, Habimana, and Hu}{Liang
  et~al\mbox{.}}{2019}]%
        {DANN}
\bibfield{author}{\bibinfo{person}{Tianan Liang}, \bibinfo{person}{Yuhua Li},
  \bibinfo{person}{Ruixuan Li}, \bibinfo{person}{Xiwu Gu},
  \bibinfo{person}{Olivier Habimana}, {and} \bibinfo{person}{Yi Hu}.}
  \bibinfo{year}{2019}\natexlab{}.
\newblock \showarticletitle{Personalizing Session-based Recommendation with
  Dual Attentive Neural Network}. In \bibinfo{booktitle}{\emph{IJCNN}}. IEEE,
  \bibinfo{pages}{1--8}.
\newblock


\bibitem[\protect\citeauthoryear{Liu, Zeng, Mokhosi, and Zhang}{Liu
  et~al\mbox{.}}{2018}]%
        {liu2018stamp}
\bibfield{author}{\bibinfo{person}{Qiao Liu}, \bibinfo{person}{Yifu Zeng},
  \bibinfo{person}{Refuoe Mokhosi}, {and} \bibinfo{person}{Haibin Zhang}.}
  \bibinfo{year}{2018}\natexlab{}.
\newblock \showarticletitle{STAMP: short-term attention/memory priority model
  for session-based recommendation}. In \bibinfo{booktitle}{\emph{KDD}}.
  \bibinfo{pages}{1831--1839}.
\newblock


\bibitem[\protect\citeauthoryear{Ludewig, Mauro, Latifi, and Jannach}{Ludewig
  et~al\mbox{.}}{2019}]%
        {ludewig2019performance}
\bibfield{author}{\bibinfo{person}{Malte Ludewig}, \bibinfo{person}{Noemi
  Mauro}, \bibinfo{person}{Sara Latifi}, {and} \bibinfo{person}{Dietmar
  Jannach}.} \bibinfo{year}{2019}\natexlab{}.
\newblock \showarticletitle{Performance comparison of neural and non-neural
  approaches to session-based recommendation}. In
  \bibinfo{booktitle}{\emph{Proceedings of the 13th ACM conference on
  recommender systems}}. \bibinfo{pages}{462--466}.
\newblock


\bibitem[\protect\citeauthoryear{Qiu, Li, Huang, and Yin}{Qiu
  et~al\mbox{.}}{2019}]%
        {fgnn}
\bibfield{author}{\bibinfo{person}{Ruihong Qiu}, \bibinfo{person}{Jingjing Li},
  \bibinfo{person}{Zi Huang}, {and} \bibinfo{person}{Hongzhi Yin}.}
  \bibinfo{year}{2019}\natexlab{}.
\newblock \showarticletitle{Rethinking the item order in session-based
  recommendation with graph neural networks}. In
  \bibinfo{booktitle}{\emph{CIKM}}. \bibinfo{pages}{579--588}.
\newblock


\bibitem[\protect\citeauthoryear{Quadrana, Karatzoglou, Hidasi, and
  Cremonesi}{Quadrana et~al\mbox{.}}{2017}]%
        {H-RNN}
\bibfield{author}{\bibinfo{person}{Massimo Quadrana},
  \bibinfo{person}{Alexandros Karatzoglou}, \bibinfo{person}{Bal{\'a}zs
  Hidasi}, {and} \bibinfo{person}{Paolo Cremonesi}.}
  \bibinfo{year}{2017}\natexlab{}.
\newblock \showarticletitle{Personalizing session-based recommendations with
  hierarchical recurrent neural networks}. In
  \bibinfo{booktitle}{\emph{proceedings of the Eleventh ACM Conference on
  Recommender Systems}}. \bibinfo{pages}{130--137}.
\newblock


\bibitem[\protect\citeauthoryear{Ren, Chen, Li, Ren, Ma, and De~Rijke}{Ren
  et~al\mbox{.}}{2019}]%
        {ren2019repeatnet}
\bibfield{author}{\bibinfo{person}{Pengjie Ren}, \bibinfo{person}{Zhumin Chen},
  \bibinfo{person}{Jing Li}, \bibinfo{person}{Zhaochun Ren},
  \bibinfo{person}{Jun Ma}, {and} \bibinfo{person}{Maarten De~Rijke}.}
  \bibinfo{year}{2019}\natexlab{}.
\newblock \showarticletitle{Repeatnet: A repeat aware neural recommendation
  machine for session-based recommendation}. In
  \bibinfo{booktitle}{\emph{AAAI}}, Vol.~\bibinfo{volume}{33}.
  \bibinfo{pages}{4806--4813}.
\newblock


\bibitem[\protect\citeauthoryear{Rendle, Freudenthaler, and
  Schmidt-Thieme}{Rendle et~al\mbox{.}}{2010}]%
        {rendle2010factorizing}
\bibfield{author}{\bibinfo{person}{Steffen Rendle}, \bibinfo{person}{Christoph
  Freudenthaler}, {and} \bibinfo{person}{Lars Schmidt-Thieme}.}
  \bibinfo{year}{2010}\natexlab{}.
\newblock \showarticletitle{Factorizing personalized markov chains for
  next-basket recommendation}. In \bibinfo{booktitle}{\emph{WWW}}.
  \bibinfo{pages}{811--820}.
\newblock


\bibitem[\protect\citeauthoryear{Sarwar, Karypis, Konstan, and Riedl}{Sarwar
  et~al\mbox{.}}{2001}]%
        {sarwar2001item}
\bibfield{author}{\bibinfo{person}{Badrul Sarwar}, \bibinfo{person}{George
  Karypis}, \bibinfo{person}{Joseph Konstan}, {and} \bibinfo{person}{John
  Riedl}.} \bibinfo{year}{2001}\natexlab{}.
\newblock \showarticletitle{Item-based collaborative filtering recommendation
  algorithms}. In \bibinfo{booktitle}{\emph{WWW}}. \bibinfo{pages}{285--295}.
\newblock


\bibitem[\protect\citeauthoryear{Schlichtkrull, Kipf, Bloem, Van Den~Berg,
  Titov, and Welling}{Schlichtkrull et~al\mbox{.}}{2018}]%
        {schlichtkrull2018modeling}
\bibfield{author}{\bibinfo{person}{Michael Schlichtkrull},
  \bibinfo{person}{Thomas~N Kipf}, \bibinfo{person}{Peter Bloem},
  \bibinfo{person}{Rianne Van Den~Berg}, \bibinfo{person}{Ivan Titov}, {and}
  \bibinfo{person}{Max Welling}.} \bibinfo{year}{2018}\natexlab{}.
\newblock \showarticletitle{Modeling relational data with graph convolutional
  networks}. In \bibinfo{booktitle}{\emph{European semantic web conference}}.
  Springer, \bibinfo{pages}{593--607}.
\newblock


\bibitem[\protect\citeauthoryear{Shani, Heckerman, Brafman, and
  Boutilier}{Shani et~al\mbox{.}}{2005}]%
        {shani2005mdp}
\bibfield{author}{\bibinfo{person}{Guy Shani}, \bibinfo{person}{David
  Heckerman}, \bibinfo{person}{Ronen~I Brafman}, {and} \bibinfo{person}{Craig
  Boutilier}.} \bibinfo{year}{2005}\natexlab{}.
\newblock \showarticletitle{An MDP-based recommender system.}
\newblock \bibinfo{journal}{\emph{Journal of Machine Learning Research}}
  \bibinfo{volume}{6}, \bibinfo{number}{9} (\bibinfo{year}{2005}).
\newblock


\bibitem[\protect\citeauthoryear{Song, Shen, Ou, Zhang, Xiao, and Liang}{Song
  et~al\mbox{.}}{2019}]%
        {song2019islf}
\bibfield{author}{\bibinfo{person}{Jing Song}, \bibinfo{person}{Hong Shen},
  \bibinfo{person}{Zijing Ou}, \bibinfo{person}{Junyi Zhang},
  \bibinfo{person}{Teng Xiao}, {and} \bibinfo{person}{Shangsong Liang}.}
  \bibinfo{year}{2019}\natexlab{}.
\newblock \showarticletitle{ISLF: Interest Shift and Latent Factors Combination
  Model for Session-based Recommendation.}. In
  \bibinfo{booktitle}{\emph{IJCAI}}. \bibinfo{pages}{5765--5771}.
\newblock


\bibitem[\protect\citeauthoryear{Tang and Wang}{Tang and Wang}{2018}]%
        {2018Personalized}
\bibfield{author}{\bibinfo{person}{Jiaxi Tang} {and} \bibinfo{person}{Ke
  Wang}.} \bibinfo{year}{2018}\natexlab{}.
\newblock \showarticletitle{Personalized Top-N Sequential Recommendation via
  Convolutional Sequence Embedding}. In \bibinfo{booktitle}{\emph{the Eleventh
  ACM International Conference}}.
\newblock


\bibitem[\protect\citeauthoryear{Wang, Cao, Wang, Sheng, Orgun, and Lian}{Wang
  et~al\mbox{.}}{2021}]%
        {wang2021survey}
\bibfield{author}{\bibinfo{person}{Shoujin Wang}, \bibinfo{person}{Longbing
  Cao}, \bibinfo{person}{Yan Wang}, \bibinfo{person}{Quan~Z Sheng},
  \bibinfo{person}{Mehmet~A Orgun}, {and} \bibinfo{person}{Defu Lian}.}
  \bibinfo{year}{2021}\natexlab{}.
\newblock \showarticletitle{A survey on session-based recommender systems}.
\newblock \bibinfo{journal}{\emph{ACM Computing Surveys (CSUR)}}
  \bibinfo{volume}{54}, \bibinfo{number}{7} (\bibinfo{year}{2021}),
  \bibinfo{pages}{1--38}.
\newblock


\bibitem[\protect\citeauthoryear{Wang, Yin, Sadiq, Chen, Xie, and Zhou}{Wang
  et~al\mbox{.}}{2016}]%
        {wang2016spore}
\bibfield{author}{\bibinfo{person}{Weiqing Wang}, \bibinfo{person}{Hongzhi
  Yin}, \bibinfo{person}{Shazia Sadiq}, \bibinfo{person}{Ling Chen},
  \bibinfo{person}{Min Xie}, {and} \bibinfo{person}{Xiaofang Zhou}.}
  \bibinfo{year}{2016}\natexlab{}.
\newblock \showarticletitle{SPORE: A sequential personalized spatial item
  recommender system}. In \bibinfo{booktitle}{\emph{ICDE}}. IEEE,
  \bibinfo{pages}{954--965}.
\newblock


\bibitem[\protect\citeauthoryear{Wang, Zhang, Liu, Liu, Zhang, Lin, and
  Zha}{Wang et~al\mbox{.}}{2020b}]%
        {wang2020beyond}
\bibfield{author}{\bibinfo{person}{Wen Wang}, \bibinfo{person}{Wei Zhang},
  \bibinfo{person}{Shukai Liu}, \bibinfo{person}{Qi Liu}, \bibinfo{person}{Bo
  Zhang}, \bibinfo{person}{Leyu Lin}, {and} \bibinfo{person}{Hongyuan Zha}.}
  \bibinfo{year}{2020}\natexlab{b}.
\newblock \showarticletitle{Beyond clicks: Modeling multi-relational item graph
  for session-based target behavior prediction}. In
  \bibinfo{booktitle}{\emph{Proceedings of The Web Conference 2020}}.
  \bibinfo{pages}{3056--3062}.
\newblock


\bibitem[\protect\citeauthoryear{Wang, Wei, Cong, Li, Mao, and Qiu}{Wang
  et~al\mbox{.}}{2020a}]%
        {gce-gnn}
\bibfield{author}{\bibinfo{person}{Ziyang Wang}, \bibinfo{person}{Wei Wei},
  \bibinfo{person}{Gao Cong}, \bibinfo{person}{Xiao-Li Li},
  \bibinfo{person}{Xian-Ling Mao}, {and} \bibinfo{person}{Minghui Qiu}.}
  \bibinfo{year}{2020}\natexlab{a}.
\newblock \showarticletitle{Global context enhanced graph neural networks for
  session-based recommendation}. In \bibinfo{booktitle}{\emph{SIGIR}}.
  \bibinfo{pages}{169--178}.
\newblock


\bibitem[\protect\citeauthoryear{Wu, Tang, Zhu, Wang, Xie, and Tan}{Wu
  et~al\mbox{.}}{2019}]%
        {srgnn}
\bibfield{author}{\bibinfo{person}{Shu Wu}, \bibinfo{person}{Yuyuan Tang},
  \bibinfo{person}{Yanqiao Zhu}, \bibinfo{person}{Liang Wang},
  \bibinfo{person}{Xing Xie}, {and} \bibinfo{person}{Tieniu Tan}.}
  \bibinfo{year}{2019}\natexlab{}.
\newblock \showarticletitle{Session-based recommendation with graph neural
  networks}. In \bibinfo{booktitle}{\emph{AAAI}}, Vol.~\bibinfo{volume}{33}.
  \bibinfo{pages}{346--353}.
\newblock


\bibitem[\protect\citeauthoryear{Xia, Yin, Yu, Wang, Cui, and Zhang}{Xia
  et~al\mbox{.}}{2020}]%
        {2020DHCN}
\bibfield{author}{\bibinfo{person}{Xin Xia}, \bibinfo{person}{Hongzhi Yin},
  \bibinfo{person}{Junliang Yu}, \bibinfo{person}{Qinyong Wang},
  \bibinfo{person}{Lizhen Cui}, {and} \bibinfo{person}{Xiangliang Zhang}.}
  \bibinfo{year}{2020}\natexlab{}.
\newblock \showarticletitle{Self-Supervised Hypergraph Convolutional Networks
  for Session-based Recommendation}.
\newblock \bibinfo{journal}{\emph{arXiv preprint arXiv:2012.06852}}
  (\bibinfo{year}{2020}).
\newblock


\bibitem[\protect\citeauthoryear{Xu, Zhao, Liu, Sheng, Xu, Zhuang, Fang, and
  Zhou}{Xu et~al\mbox{.}}{2019}]%
        {xu2019graph}
\bibfield{author}{\bibinfo{person}{Chengfeng Xu}, \bibinfo{person}{Pengpeng
  Zhao}, \bibinfo{person}{Yanchi Liu}, \bibinfo{person}{Victor~S Sheng},
  \bibinfo{person}{Jiajie Xu}, \bibinfo{person}{Fuzhen Zhuang},
  \bibinfo{person}{Junhua Fang}, {and} \bibinfo{person}{Xiaofang Zhou}.}
  \bibinfo{year}{2019}\natexlab{}.
\newblock \showarticletitle{Graph Contextualized Self-Attention Network for
  Session-based Recommendation.}. In \bibinfo{booktitle}{\emph{IJCAI}},
  Vol.~\bibinfo{volume}{19}. \bibinfo{pages}{3940--3946}.
\newblock


\bibitem[\protect\citeauthoryear{Xu, Wu, Wang, Feng, Witbrock, and Sheinin}{Xu
  et~al\mbox{.}}{2018}]%
        {xu2018graph2seq}
\bibfield{author}{\bibinfo{person}{Kun Xu}, \bibinfo{person}{Lingfei Wu},
  \bibinfo{person}{Zhiguo Wang}, \bibinfo{person}{Yansong Feng},
  \bibinfo{person}{Michael Witbrock}, {and} \bibinfo{person}{Vadim Sheinin}.}
  \bibinfo{year}{2018}\natexlab{}.
\newblock \showarticletitle{Graph2seq: Graph to sequence learning with
  attention-based neural networks}.
\newblock \bibinfo{journal}{\emph{arXiv preprint arXiv:1804.00823}}
  (\bibinfo{year}{2018}).
\newblock


\bibitem[\protect\citeauthoryear{You, Wang, Pal, Eksombatchai, Rosenburg, and
  Leskovec}{You et~al\mbox{.}}{2019}]%
        {HTCN}
\bibfield{author}{\bibinfo{person}{Jiaxuan You}, \bibinfo{person}{Yichen Wang},
  \bibinfo{person}{Aditya Pal}, \bibinfo{person}{Pong Eksombatchai},
  \bibinfo{person}{Chuck Rosenburg}, {and} \bibinfo{person}{Jure Leskovec}.}
  \bibinfo{year}{2019}\natexlab{}.
\newblock \showarticletitle{Hierarchical temporal convolutional networks for
  dynamic recommender systems}. In \bibinfo{booktitle}{\emph{The world wide web
  conference}}. \bibinfo{pages}{2236--2246}.
\newblock


\bibitem[\protect\citeauthoryear{Zhang, Wu, Gao, Jiang, Xu, and Wang}{Zhang
  et~al\mbox{.}}{2020}]%
        {A-PGNN}
\bibfield{author}{\bibinfo{person}{Mengqi Zhang}, \bibinfo{person}{Shu Wu},
  \bibinfo{person}{Meng Gao}, \bibinfo{person}{Xin Jiang}, \bibinfo{person}{Ke
  Xu}, {and} \bibinfo{person}{Liang Wang}.} \bibinfo{year}{2020}\natexlab{}.
\newblock \showarticletitle{Personalized graph neural networks with attention
  mechanism for session-aware recommendation}.
\newblock \bibinfo{journal}{\emph{IEEE Transactions on Knowledge and Data
  Engineering}} (\bibinfo{year}{2020}).
\newblock


\end{thebibliography}

% \appendix

% \input{appendix.tex}

\end{document}